\documentclass{emulateapj}

\newcommand{\kms}{\ensuremath{{\rm km~s}^{-1}}}

\newcommand{\vmax}{v_\mathrm{max}}

\shorttitle{Physical Properties of Quasars and Outflows}
\shortauthors{Ganguly et al.}

\begin{document}

\title{Outflows and the Physical Properties of Quasars}

\author{Rajib Ganguly\altaffilmark{1},
Michael S. Brotherton\altaffilmark{1}, Sabrina
Cales\altaffilmark{1}, Brian Scoggins\altaffilmark{1}, Zhaohui
Shang\altaffilmark{1,2}, Marianne Vestergaard\altaffilmark{3}}

\altaffiltext{1}{Department of Physics and Astronomy, University of
Wyoming (Dept. 3905), 1000 East University Ave., Laramie, WY 82072}

\altaffiltext{2}{Department of Physics, Tianjin Normal University,
300074 Tianjin, China}

\altaffiltext{3}{Department of Astronomy and Steward Observatory 933
N Cherry Ave., Tucson AZ 85721-0065}

\begin{abstract}

We have investigated a sample of 5088 quasars from the Sloan Digital
Sky Survey Second Data Release in order to determine how the
frequency and properties of broad absorptions lines (BALs) depend on
black hole mass, bolometric luminosity, Eddington fraction
($L/L_{Edd}$), and spectral slope.  We focus only on high-ionization
BALs and find a number of significant results. While quasars
accreting near the Eddington limit are more likely to show BALs than
lower $L/L_{Edd}$ systems, BALs are present in quasars accreting at
only a few percent Eddington.  We find a stronger effect with
bolometric luminosity, such that the most luminous quasars are more
likely to show BALs. There is an additional effect, previously
known, that BAL quasars are redder on average than unabsorbed
quasars. The strongest effects involving the quasar physical
properties and BAL properties are related to terminal outflow
velocity.  Maximum observed outflow velocities increase with both
the bolometric luminosity and the blueness of the spectral slope,
suggesting that the ultraviolet luminosity to a great extent
determines the acceleration.  These results support the idea of
outflow acceleration via ultraviolet line scattering.

\end{abstract}

\keywords{quasars: general --- quasars: absorption lines ---
quasars: fundamental parameters}

\section{Introduction}

The nature and significance of broad absorption lines (hereafter,
BALs) seen in some 10-20\% of high luminosity quasars is still not
apparent. The broad, blueshifted lines indicate that high velocity
outflows are present in at least some quasars.  The outflows are
potentially important as they may enable accretion through carrying
away angular momentum, and they may significantly chemically enrich
the interstellar medium of the quasar host galaxy and the
surrounding intergalactic medium.

There are a number of potential explanations for observed frequency
of BALs. In the interest of unified schemes, it is postulated that
all high-luminosity quasars host the outflows that give rise to
BALs. This postulate is not without warrant, as quasars with and
without BALs appear to show similar intrinsic observed continuum and
emission-line properties \citep[e.g.,][]{weymann91,gallsc}. In its
simplest interpretation, the frequency of observed BALs is tied to
the covering factor of the outflow around the central black hole.
That is, whether or not one observes the outflow in absorption
depends on the orientation of the central engine
\citep[e.g.,][]{weymann91,goodrich97,kv98}. Early evidence from
spectro-polarimetry of BALs bolstered the idea that BAL quasars were
viewed edge-on (or nearly so) and the outflow was equatorial
\citep[e.g.,][]{good95}. This explanation is difficult to reconcile
in the face of new evidence from the studies of radio-loud BALs. The
discovery of radio-loud BAL quasars \citep[e.g.,][]{becker00} opens
up the possibility of directly gauging the orientations of BALs.
Mounting evidence in the form of spectro-polarimetry and brightness
temperatures \citep[e.g.,][and references therein]{brotherton06} is
showing that BALs are viewed at a large variety of viewing angles
(from $\sim15^\circ$\ from the jet axis to nearly edge-on).

Two plausible alternative explanations include unification in the
time-domain or more complicated orientation schemes. The former case
purports that BALs are a short-duration (possibly episodic) phase in
the duty cycle of the accreting black hole
\citep*[e.g.,][]{vwk93,becker00,gregg00,gbd06}. Such a scenario also
tends to connect BALs to even rarer objects like post-startburst
quasars \citep[e.g.,][]{bro99} and to more extreme objects
ultra-luminous infrared galaxies \citep[e.g.,][]{sanders88} in
suggestive evolutionary sequences.

In the latter case, the geometry of the outflow, and hence the
frequency with which it is intercepted producing a BAL, is dependent
on the intrinsic SED (or physical parameters) of the disk. For
instance, in the conventional wisdom, the physical parameters that
are most strongly tied to the presence of a BAL outflow is the
Eddington ratio and the black hole mass
\citep[e.g.,][]{bg92,boroson02,pk04}. \citet{boroson02} and
\citet{yw03} have shown that BAL quasars have strong \ion{Fe}{2} and
weak [\ion{O}{3}] emission putting them at one extreme of the
\citet{bg92} Eigenvector 1, which is thought to be driven to the
accretion rate in Eddington units. In this scheme, BAL quasars are
thought to be the more massive (hence, more luminous) analogs of
narrow-line Seyfert 1 galaxies \citep[e.g.,][]{bg00,boroson02}.

After many decades of work, we now have the means to reliably
estimate the fundamental physical properties (e.g., black hole mass,
bolometric luminosity, Eddington ratio) of a quasar. Thus, we are in
a position to ask whether (and how) the parameters of a BAL outflow
depend on the physical quasar properties. With the large numbers of
quasars available through the Sloan Digital Sky Survey, we can do so
in a highly statistically significant manner, where only systematic
uncertainties affect our results. Our approach here is to address
and answer some of these questions about BAL quasars by determining
their fundamental physical properties and comparing them to normal
quasars (i.e., to make differential comparisons where systematics
should not affect the result) and to look for correlations with BAL
properties. We explain the details of our methodology in
\S\ref{sec:datamethod}, our basic results in \S\ref{sec:analysis},
discuss the results in \S\ref{sec:discussion}, and summarize our
conclusions in \S\ref{sec:summary}. We adopt a cosmology with
$\Omega_M = 0.3$, $\Omega_\Lambda = 0.7$, $H_0$ = 70 km s$^{-1}$
Mpc$^{-1}$ throughout this paper.

\section{Data and Methods}
\label{sec:datamethod}
\subsection{Selection and Classification}

Our sample comes from the Sloan Digital Sky Survey
\citep[SDSS;][]{york00}, Data Release 2 \citep[DR2;][]{sdss2}. We
requested all 5088 objects classified as having broad emission
lines, i.e., quasars, with redshifts between 1.7 and 2.0. This
redshift range places both the \ion{C}{4} $\lambda$1549 and
\ion{Mg}{2} $\lambda$2800 emission lines in the window of the SDSS
spectra (which covers the range $\sim$3820--9200\,\AA). The
\ion{C}{4} region permits us to identify BALs, while the \ion{Mg}{2}
region can be used to make virial mass estimates \citep{mj02}. The
BAL absorption associated with \ion{C}{4} makes that line a bad
choice for making mass estimates. When BALs are found associated
with low-ionization species like \ion{Al}{3} $\lambda$1860, and
\ion{Mg}{2}, then the \ion{Mg}{2} line is also compromised much of
the time. This measurement is required for black hole mass
estimation, which in turn is needed to determine the Eddington
luminosity.  In addition, low-ionization BAL quasars appear
systematically reddened compared to other quasar classes
\citep[e.g.,][]{sf92,yv99,becker00,ndb00,bro01,hallai,gtr03,reichard03b},
making the determination of the continuum luminosity more uncertain
and probably biased unless an uncertain correction is made.
Therefore, we have excluded low-ionization BAL quasars from our
subsequent analysis (leaving 5033 total objects). We make an
additional cut below when we fit the \ion{Mg}{2} region of the
spectra.

\begin{deluxetable}{lrcrc}
\tablewidth{0pc}
\tablecaption{Classification of SDSS-DR2 $1.7 \leq z \leq 2$\
Quasars}
\tablehead {
& \multicolumn{2}{c}{Pre-Fitting} & \multicolumn{2}{c}{Post-Fitting} \\
& \multicolumn{2}{c}{\hrulefill} & \multicolumn{2}{c}{\hrulefill} \\
\colhead{Class} & \colhead{Number} & \colhead{Fraction} & \colhead{Number} & \colhead{Fraction}}
\startdata
LoBALs         &   55 &   1\%   & \nodata &        \\
mini-BALs/BALs &  562 &  11.0\% &     536 & 10.5\% \\
AALs           & 1898 &  37.3\% &    1813 & 35.6\% \\
Unabsorbed     & 2573 &  50.6\% &    2509 & 49.3\% \\ \hline \\[-7pt]
Total          & 5088 &         &    4858 & 95.5\%
\tablecomments{The columns on the left indicate our subjective classifications
before applying our fitting prescriptions to measure the \ion{Mg}{2}\ emission line
region. The columns on the right indicate the number of quasars that had reasonable fits.
Bad fits were typically due to poor S/N, or the presence of intervening or associated
absorption contaminating the \ion{Mg}{2}\ emission line.}
\enddata
\label{tab:demo}
\end{deluxetable}

We subjectively classify objects (i.e., through visual inspection)
into three classes: (1) objects displaying clear signs of an outflow
(mini-BALs/BALs); (2) objects showing no signs of intrinsic
absorption (Unabsorbed); and (3) objects that have absorption near
the \ion{C}{4} emission line that may break up into discrete
components at higher resolution (AALs\footnote{The AAL abbreviation
stands for ``associated'' absorption lines. Nominally, these are
lines that have a narrow velocity-dispersion ($\lesssim500$\,\kms)
and appear near the quasar redshift ($c|\Delta z| \lesssim
5000$\,\kms). We do not adhere strictly to these criteria in our
subjective classification scheme. {C\,{\sc iv}} absorption that is
narrow or sufficiently clumpy and appears superposed on the {C\,{\sc
iv}} emission line is deemed an AAL.}). The incidence of each of
these classes in listed in the second and third columns of
Table~\ref{tab:demo}. In our subjective scheme, we adhere to the
idea that an outflow seen in absorption should be relatively smooth
(i.e., that the profile is unlikely to break up into more discrete
components if observed at higher dispersion). If a profile appears
too clumpy (i.e., consists of narrow components), then it is placed
in the AAL class, if the absorption takes place near the emission
line. If the absorption has a narrow velocity dispersion (FWHM
$\lesssim$500\,\kms), is clumpy and appears at a large blueshift, it
is taken to result from intervening structures (which typically is
also accompanied by narrow low-ionization absorption lines). Since
there is likely a continuum of velocity widths that arise from
absorption by outflows, there will be a cases where our
classification is incorrect. To combat this, four of the authors
(Ganguly, Brotherton, Cales, and Scoggins) have independently
classified the spectra. Comparison between the authors leads to very
few cases where there is any dispute, implying that our
classifications are both uniform and reproducible. Using only a
single-epoch low-dispersion spectrum, we feel that this is the best
that can be accomplished without resorting to more quantitative
schemes. [Such schemes would require continuum-fitting procedures,
such as template fitting or polynomial fitting, and are beyond the
scope of this investigation.]

While our classifications are subjective, we feel they are more
complete than the various quantitative schemes from the literature
which are either insufficient, biased, or too contaminated with
false-positives for our purposes. The BALnicity Index (BI) was
defined by \citet{weymann91} in order to be sure they classified
only BAL quasars as BAL quasars with low signal-to-noise ratio and
low-resolution LBQS spectra. For example, intrinsic absorption
systems that appear at high velocity (i.e., mini-BALs) can be
excluded when using a BI criterion. Intrinsic absorption appearing
within 5000\,\kms\ would also be excluded by a BI selection.
Similarly, the Absorption Index (AI) of \citet{hallai}\footnote{See
\citet{trump06} for a revised definition of this parameter.}, while
more liberal than the BI, still has some fairly arbitrary limits to
help weed out blends of associated absorbers that may be intervening
systems. Since we wish to test the properties of outflows (e.g.,
maximum velocity of absorption, onset velocity, velocity width) as a
function of other quasar properties (e.g., luminosity, black hole
mass), it is important that our sample be roughly complete. We
emphasize here that absorption by outflows comes from a continuum of
velocity widths. While a small error rate in our classifications
would not compromise our study, inclusion of only BI$>$0, or AI$>$0
objects in a sample of objects with outflows would bias our results.

We compare our subjectively-selected sample to the more objective
BAL quasar sample from the SDSS third data release compiled by
\citet{trump06}. In total, we find that 562 of the 5088 objects in
our sample appear to have spectroscopic evidence of high-ionization
outflowing gas (i.e., with no accompanying low-ionization
absorption). By comparison, 1206 of the 5088 objects appear in the
\citet{trump06} BAL catalog with the following classifications: Hi
-- 835, nHi -- 321, H -- 2, Lo -- 41, nLo -- 3, LoF -- 3, nLoF -- 1.
[A ``Hi'' or ``H'' classification indicates the presence of
\ion{C}{4}\ absorption, while a ``Lo'' classification indicates the
additional presence of \ion{Mg}{2}\ absorption. An ``n'' prefix
indicates that the absorption width is narrow while a ``F''
indicates the presence of \ion{Fe}{2}\ absorption.] That is,
\citet{trump06} appear to find 1158 objects with evidence for
high-ionization outflows. At face value, it would seem that our
subjective classification scheme is not more efficient at selecting
outflows. We present a cross-comparison of our classification scheme
and that of \citet{trump06} in Table~\ref{tab:trump}. In the first
two rows of the table, we present a head-to-head comparison of the
total number objects in our work and those objects that would have
been flagged as BALs by \citet{trump06}, as well as the breakdown
with our classifications. In the next two rows (rows three and
four), we further break down the \citet{trump06} numbers by
ionization. In the remaining rows of the table, we present the full
demographics using the \citet{trump06} classification and their
comparison with our classification scheme. We draw the reader's
attention to the following comparisons:
\begin{deluxetable}{lrrrrr}
\tablewidth{0pc}
\tablecaption{Comparison to \citet{trump06} Classification Scheme}
\tablehead {
& \colhead{Total} & \colhead{Unabs.} & \colhead{AALs} & \colhead{BAL} & \colhead{LoBAL}
}
\startdata
This Work                  & 5088 & 2572 & 1898 & 562 & 55 \\
T06, AI$>0$    & 1206 &  181 &  420 & 551 & 54 \\
T06, AI$\leq0$ & 3882 & 2391 & 1478 &  11 &  1 \\ \hline \\[-7pt]
\multicolumn{6}{c}{Breakdown of \citet{trump06} AI$>0$\ classes} \\ \hline \\[-7pt]
high-ionization            & 1158 & 178  & 400  & 540 & 40 \\
low-ionization             & 48   & 3    & 20   & 11  & 14 \\ \hline
Hi                         & 835  & 72   & 186  & 537 & 40 \\
nHi                        & 321  & 106  & 214  & 1   & 0  \\
H                          & 2    & 0    & 0    & 2   & 0  \\
Lo                         & 41   & 3    & 16   & 10  & 12 \\
nLo                        & 3    & 0    & 1    & 1   & 1  \\
LoF                        & 3    & 0    & 2    & 0   & 1  \\
nLoF                       & 1    & 0    & 1    & 0   & 0
\enddata
\tablecomments{T06 indicates \citet{trump06}. The fourth row, labeled ``high-ionization,'' denotes the sum of the
``Hi,'' ``nHi,'' and ``H'' classifications. The fifth row, labeled ``low-ionization,''
denotes the sum of the ``Lo,'' ``nLo,'' ``LoF,'' and ``nLoF'' classifications. See
text for a description of the sub-classes.}
\label{tab:trump}
\end{deluxetable}

\begin{figure*}
\begin{center}
\epsscale{0.7} \rotatebox{-90}{\plotone{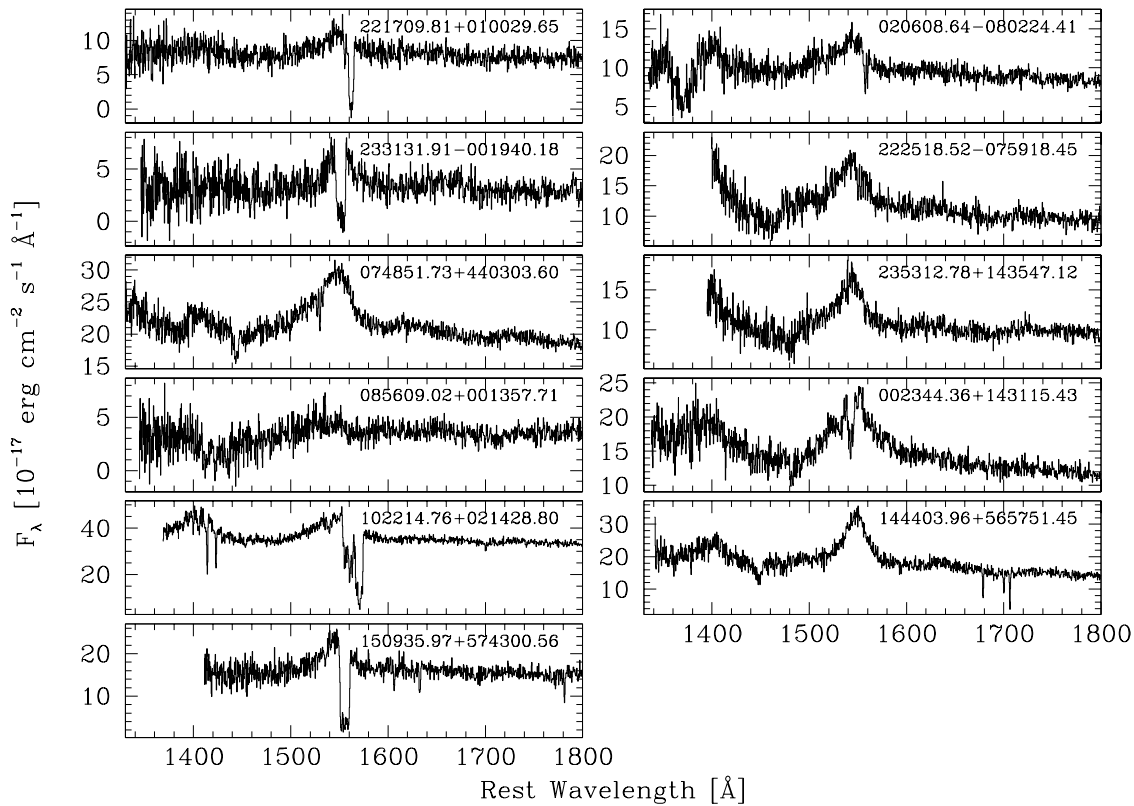}} \caption{We show
spectra of the 11 objects that we subjectively classify as HiBALs,
that do not appear in the \citet{trump06} catalog.}
\label{fig:montage1}
\end{center}
\end{figure*}

\begin{figure}
\epsscale{0.7}
\plotone{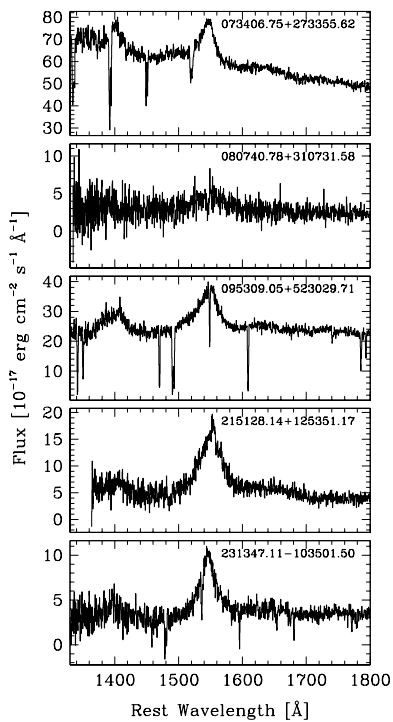}
\caption{We show example spectra of objects selected by
\citet{trump06} as HiBALs which do not appear to show intrinsic
absorption by our subjective classification.} \label{fig:montage2}
\end{figure}

\begin{itemize}
\item[1.] Of the 562 quasars with high-ionization outflows (BALs) that
we selected, \citet{trump06} only cataloged 551 objects. Of these,
only 540 have a ``Hi,'' ``nHi,'' or ``H'' classification. The other
11/551 objects (that are in the catalog) are given ``Lo'' or ``nLo''
classifications. The 11/562 objects that are not in the
\citet{trump06} catalog are shown in Figure~\ref{fig:montage1}. Of
these, six are cases where either the velocity limits of integration
for AI or the continuous absorption criterion are not sufficient. In
other cases, the profile may be too shallow relative to the
signal-to-noise to allow a precise AI measurement.
\item[2.] 181 of the objects classified by \citet{trump06} as
a BAL appear to be intrinsically unabsorbed objects, typically with
\ion{Mg}{2}\ or \ion{Fe}{2}\ absorption by intervening structures.
Examples of these objects are shown in Figure~\ref{fig:montage2}.
\item[3.] 40 of the objects that we classify as LoBALs are presented as
HiBALs in the \citet{trump06} catalog. These objects appear either
somewhat reddened or have broad absorption near the \ion{Mg}{2}\ or
\ion{Al}{3} emission lines, which may affect our ability to carry
out spectral fits of the \ion{Mg}{2} region.
\item[4.] For 400 of the objects classified by \citet{trump06} as
high-ionization BAL quasars (``Hi'' or ``nHi''), it is not clear
that the \ion{C}{4} absorption is necessarily a result of an
outflow. In these cases, the absorption profiles seem as if they
would break into more discrete components if observed at higher
resolving power. The gas may be due to absorption by the host
galaxy, or other structures related to the quasar/quasar
environment. We place these objects in our AAL class.
\end{itemize}
We also note that 1478 additional quasars in our sample have AALs,
that is narrow velocity-dispersion systems appearing within
5000\,\kms\ of the quasar redshift. Some of these are likely to be
intrinsic to the quasar central engine in some form. [Estimates
range from $\geq 20$\% \citep[][]{wise04} based on time-variability
to $\sim33$\% based on partial coverage \citep[][]{misawa07}.] None
of these systems were found with the revised AI-based selection
scheme of \citet{trump06}.

We conclude that, in this sample, the number of objects with
outflows detectable in absorption is probably closer to our value of
562 (11\%), and has likely been overestimated (through the inclusion
of ``false positives'') in the \citet{trump06} catalog. While the
efforts at quantitatively selecting outflows are undergoing revision
(and have certainly progressed from the days of BALnicity), further
revisions are required to decrease the number of false-positives. If
our assessments are taken to be a truer reflection of outflow
classifications, then we estimate that the \citet{trump06} catalog
is 98\% (605/617) complete toward finding HiBALs and LoBALs
(consistent with their estimation), but suffers a 15\% (181/1206)
rate of false-positives. Detailed scrutiny of subsamples of sources,
such as presented here, should help in that goal.

\subsection{Spectral Fitting and Parameter Estimation}

Since one of our goals is to compare the physical properties (e.g.,
continuum luminosity, spectral shape, virial mass, Eddington ratio)
of absorbed and unabsorbed quasars, we carried out fits around the
\ion{Mg}{2} $\lambda2800$\ emission line (over the rest-frame
wavelength range 2000--3000\,\AA) using the {\sc specfit} task
\citep{specfit}. Our fits include a power-law continuum (with the
convention $F_\lambda \sim \lambda^{-\alpha}$), the I Zw 1 iron
emission template \citep{vw01}, and two Gaussians (a broad and a
narrow component) for the \ion{Mg}{2} emission line. The fits were
carried out on spectra that were reduced to rest-frame wavelengths
calculated using the redshift reported by SDSS.

Initially, we tried to fit all spectra with a single fitting
prescription (as summarized in column 3 of Table~\ref{tab:fits}). We
found that there was sufficient variety in the spectra that one
prescription was not adequate, and we turned to five different
prescriptions. These five prescriptions are also summarized in
Table~\ref{tab:fits}, which includes allowed ranges in the
parameters, and parameters that were tied together. For example, in
one prescription, we tied the wavelength shifts of the broad and
narrow \ion{Mg}{2} components and the \ion{Fe}{2} together (i.e., no
relative shift between them). In another prescription, all three
shifts were allowed to vary independently. We also tried
prescriptions with different ranges in various parameters like the
magnitude of the shifts or the widths of the emission line
components.

\begin{deluxetable*}{lllllll}
\tablewidth{0pc}
\tablecaption{Summary of Fitting Prescriptions}
\tablehead {
& & \multicolumn{5}{c}{Run Number} \\
Property & Unit & \colhead{1} & \colhead{2} & \colhead{3} & \colhead{4} & \colhead{5}
}
\startdata
\multicolumn{6}{l}{Power-law} \\
~~Index\tablenotemark{a}         & &     [1,10] &     [0.1,10] &      [0.1,3] &      [0.1,3] &      [0.1,3] \\
~~Normalization\tablenotemark{b} & &  [1,10000] &  [0.1,10000] &  [0.1,10000] &  [0.1,10000] &  [0.1,10000]\\ \hline \\[-7pt]
\multicolumn{6}{l}{Fe template} \\
~~Scaling          &             &   [0.001,10] &   [0.001,10] &   [0.001,10] &   [0.001,10] &   [0.001,10] \\
~~Wavelength shift & \AA         &            0 &          M2B &     [-50,50] &            0 &     [-50,50] \\
~~Width            & \kms        &  [900,10000] &  [900,12000] &  [900,12000] &  [900,12000] &  [900,12000] \\ \hline \\[-7pt]
\multicolumn{6}{l}{\ion{Mg}{2} broad component} \\
~~Scaling          &             &     [0.1,10] &    [0,30000] &    [0,30000] &    [0,30000] &    [0,30000] \\
~~Wavelength shift & \AA         &     [-28,22] &     [-28,22] &     [-18,22] &     [-18,22] &    [-98,102] \\
~~Width            & \kms        & [2000,20000] & [2000,20000] & [2000,20000] & [2000,20000] & [2000,20000] \\ \hline \\[-7pt]
\multicolumn{6}{l}{\ion{Mg}{2} narrow component} \\
~~Scaling          &             &     [0.1,10] &    [0,30000] &    [0,30000] &    [0,30000] &    [0,30000] \\
~~Wavelength shift & \AA         &     [-18,22] &          M2B &          M2B &          M2B &          M2B \\
~~Width            & \kms        &  [900,10000] &  [900,10000] &  [900,10000] &  [900,10000] &  [900,10000] \\ \hline \\[-7pt]
\multicolumn{2}{r}{Number of times chosen:} & 711 & 1687 & 1693 & 735 & 32 \\
\multicolumn{2}{r}{mini-BALs/BALs:}         &  65 &  141 &  233 &  92 &  5 \\
\multicolumn{2}{r}{AALs:}                   & 269 &  644 &  609 & 271 & 20 \\
\multicolumn{2}{r}{Unabsorbed:}             & 377 &  902 &  851 & 372 &  7
\enddata
\tablecomments{In all cases, the fits were carried out on the
rest-frame wavelength range 2000--3000\,\AA. M2B: The parameter was
tied to the same value as the corresponding parameter of the
\ion{Mg}{2} broad component.} \tablenotetext{a}{The sign convention
for the power-law index, $\alpha$, is $F_\lambda \sim
\lambda^{-\alpha}$.} \tablenotetext{b}{Units for the power law are
$10^{-17}$\ erg cm$^{-2}$ s$^{-1}$ \AA$^{-1}$} \label{tab:fits}
\end{deluxetable*}

We took the approach of fitting each spectrum using the all five
prescriptions and adopting the model that provided the best fit
(i.e., the lowest $\chi^2$\ value). The bottom four rows of the
table report the number of times each prescription was adopted for
all quasars and for the three quasar classes (mini-BALs/BALs, AALs,
Unabsorbed). In many cases, more than one prescription provides a
good fit, with similar best-fit values and the actual difference in
$\chi^2$\ values is small. In all, we were able to achieve good fits
for about 95\% of the quasar spectra. 230 of the quasars could not
be fit well due to poor signal-to-noise, or the presence of
intervening or associated absorption contaminating the \ion{Mg}{2}
emission line. In the right two columns of Table~\ref{tab:demo}, we
report how many quasars in each classification were successfully fit
with one of our prescriptions. Comparison of these numbers with the
pre-fitting numbers indicates that no single class was affected more
than the others.

\begin{figure}
\epsscale{1.0} \plotone{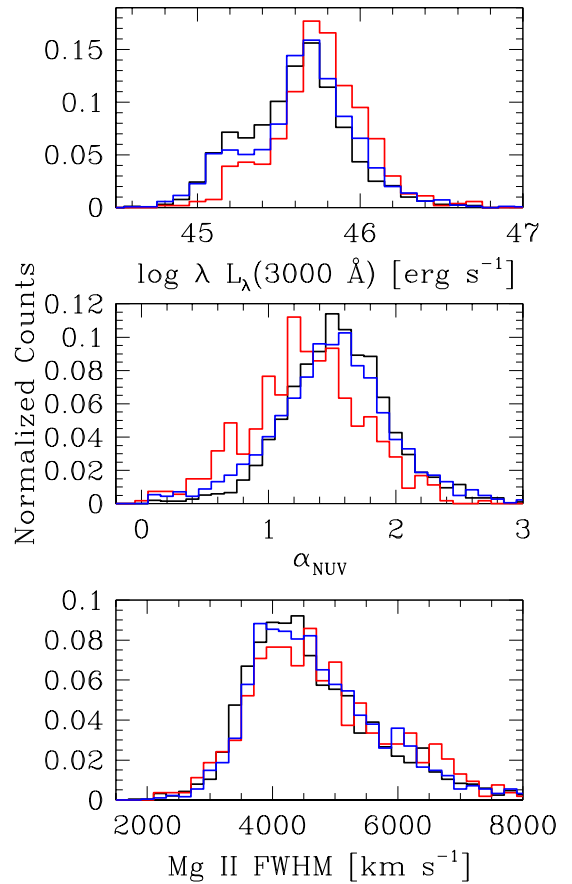}
\caption[Fitting Comparison]{We compare the continuum fitting
results between BAL QSOs (red or dark histogram), Unabsorbed QSOs
(black or shaded histogram), and AAL QSOs (blue or light histogram).
The distributions have been normalized to unit area.}
\label{fig:balregcomp}
\end{figure}

We show in Figure~\ref{fig:balregcomp} a comparison of fitted
properties (near UV spectral index, 3000\,\AA\ luminosity, and
\ion{Mg}{2} $\lambda2800$\ emission line FWHM) between three of our
quasar classes, unabsorbed quasars, BAL quasars, and AAL quasars. We
numerically computed the \ion{Mg}{2} $\lambda2800$\ emission line
FHWM from the sum of the two Gaussian components. Generally there is
good agreement between the samples with, perhaps, the mini-BAL/BAL
class having systematically broader \ion{Mg}{2} $\lambda$2800
emission-lines, flatter spectral indices, and higher 3000\,\AA\
luminosities.

From the fitting parameters, we estimate two fundamental physical
properties for all the quasars, the black hole mass, and the
Eddington ratio. For an estimation of the black hole mass for each
quasar, we use the following prescription derived by \citet{mj02}
which uses the 3000\,\AA\ monochromatic luminosity and the
\ion{Mg}{2} $\lambda$2800 FWHM:
\begin{equation}
{{M_{\mathrm{BH}}} \over {M_\odot}} = 3.37
\left ( {{\lambda L_\lambda(3000\,\mathrm{\AA})} \over {10^{42} \mathrm{erg~s}^{-1}}} \right )^{0.47}
\left [ {{\mathrm{FWHM(MgII)}} \over {\mathrm{\kms}}} \right ]^2
\end{equation}
\citet{dm04}\ note that this scaling law gives about a factor of
five smaller mass than estimates based on the \ion{C}{4} or
H$\beta$\ emission line widths \citep[e.g.,][]{kaspi00}. However,
since our primary goal is to look at {\it differences} in quasar
properties between BALs and non-BALs, this should not adversely
affect our results.

We estimate the bolometric luminosity identically for each quasar.
The measured rest-frame flux at 3000\,\AA\ from the fit to the
power-law continuum is converted into an emitted luminosity using
luminosity distances for our cosmology. We convert these
monochromatic luminosities into bolometric luminosities:
\begin{equation}
L_\mathrm{bol} = 4 \pi D_\mathrm{L}^2 f (1+z) \lambda F_\lambda,
\end{equation}
where $\lambda=3000$\,\AA, $F_\lambda$\ is the observed flux density
at 3000\,\AA, $D_\mathrm{L}$\ is the luminosity distance, and $f=5$\
is the average bolometric correction from 3000\,\AA\ \citep[see
Figure 12 from][]{gtr06}. We acknowledge here that other, and
perhaps more refined, estimates of the bolometric correction
\citep[e.g.,][]{ves04,shang05}. Use of these estimates would
generally require an extrapolation of our power-law fit to another
wavelength, which we wish to avoid. Since we are making a
differential comparison between BAL and non-BAL quasars, this should
not affect our results. \citet{gtr06} estimate $\sim$20\%
uncertainty in using a mean bolometric correction. This would have
the effect of smearing out the distribution. In the absence of yet
more refined approaches to making bolometric corrections {\it en
masse}, we feel this is best that can be done.

We calculate the Eddington luminosity for each quasar given our
estimate of the mass of its black hole: $L_\mathrm{Edd} = 1.51
\times 10^{38} (M/M_\odot)$\ erg s$^{-1}$\ \citep[e.g.,][eq.
6.21]{krolikagn}. We then compute the Eddington ratio,
$L_\mathrm{bol}/L_\mathrm{Edd}$. In Table~\ref{tab:qsos}, we present
our measurements of these quasars, as well as the two derived
physical parameters (black hole mass, Eddington ratio), the fitting
run that provided the best fit, and our classification.

\begin{deluxetable}{lrrr}
\tablewidth{0pc}
\tablecaption{Outflow Absorption-Line Measurements}
\tablehead {
\colhead{Target} & \colhead{BI}     & \colhead{$v_\mathrm{min}$} & \colhead{$\vmax$} \\
                 & \colhead{(\kms)} & \colhead{(\kms)}           & \colhead{(\kms)}
}
\startdata
SDSS J002344.36$+$143115.43 &   0 &  6964 & 27697 \\
SDSS J020608.64$-$080224.41 &   0 & 32771 & 39285 \\
SDSS J074851.73$+$440303.60 &   0 & 15421 & 25505 \\
SDSS J085609.02$+$001357.71 &   0 & 16512 & 28419 \\
SDSS J102214.76$+$021428.80 &   0 & -4933 &   587 \\
SDSS J150935.97$+$574300.56 &   0 & -2107 &  -103 \\
SDSS J144403.96$+$565751.45 &   0 & 19375 & 23666 \\
SDSS J220900.66$-$001413.23 & 425 &  1469 &  6677 \\
SDSS J222518.52$-$075918.45 &   0 & 15453 & 24780 \\
SDSS J233131.91$-$001940.18 &   0 & -2090 &   193 \\
SDSS J235312.78$+$143547.12 &   0 &  1022 & 26929
\enddata
\label{tab:balmeas}
\end{deluxetable}

In addition to measurements of the continuum and emission-line
properties of the quasar, we also wish to test how BAL properties
depend on quasar properties. The \citet{trump06} catalog already
presents various measurements for most (but not all) of our BALs. We
adopt the BALnicity index (BI), absorption index (AI), and maximum
velocity of absorption ($\vmax$) measurements from \citet{trump06}
for the 551 BALs that appear in their catalog. For the remaining 11
BAL quasars that do not appear in the \citet{trump06} catalog, we
use our measurements of BI and $\vmax$. These are listed in
Table~\ref{tab:balmeas} (as well as the onset velocity of
absorption, $v_\mathrm{min}$). We did not measure AI values for the
11 BAL quasars because the new definition of AI to reflect a true
equivalent width requires knowledge of the (uncertain) shape of the
\ion{C}{4} emission-line since the integration range starts at zero
velocity. \citet{trump06} use a sophisticated template-fitting
algorithm to reproduce the emission-line shapes, but this is beyond
of the scope of this paper. A comparison of our BI and $\vmax$\
measurement methods to those of \citet{trump06} for a subsample of
BALs indicates that the two are consistent.

\section{Analysis}
\label{sec:analysis}
\subsection{Differences Between BAL and Non-BAL Quasars}

With measurements/derivations of the quasar physical properties, we
can explore if there are systematic differences between unabsorbed
quasars and BAL quasars. The distributions of measured properties
(Figure~\ref{fig:balregcomp}) do not show appreciable differences in
either the general shape or in the mean values
(Table~\ref{tab:means}). BAL quasars may be systematically more
luminous (at 3000\,\AA) and have redder NUV spectra ($\langle
\alpha(\mathrm{BAL}) \rangle = 1.29$, $\langle
\alpha(\mathrm{Unabs.}) \rangle = 1.55$, with standard deviations of
$\sim 0.4$), but only slightly . We note here that the absolute
shape of the luminosity distribution is affected by the SDSS quasar
selection criteria. The important observation here is that there
does not appear to be qualitatively significant {\it differences}
between the three quasar classes (Unabsorbed, AALs, and BALs); they
are all within a standard deviation.

\begin{figure}
\epsscale{1.0} \plotone{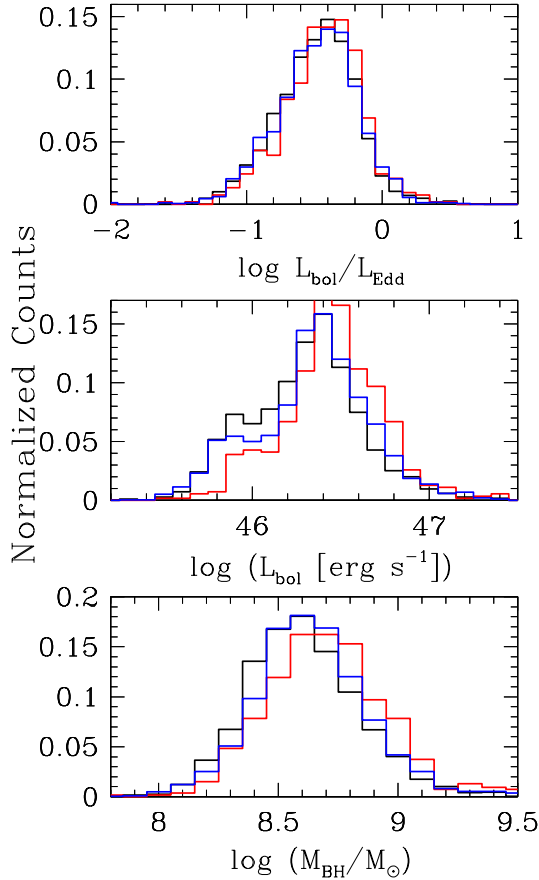}
\caption[Comparison of Derived Parameters]{We compare the
distribution of Eddington ratio (top), bolometric luminosity
(middle), and black hole mass (bottom) between BAL QSOs (red or dark
histogram), Unabsorbed QSOs (black or shaded histogram), and AAL
QSOs (blue or light histogram). The distributions have been
normalized to unit area.} \label{fig:balregcomp2}
\end{figure}

Quantitative tests (like the Kolmogorov-Smirnov test) do show that
the small differences in the distributions are apparently
significant. However, when dealing with such large samples, even
small differences in the distributions can be manifest as having
high statistical significance. We stress here that visual inspection
of Fig.~\ref{fig:balregcomp}-\ref{fig:balregcomp2} show very clear
overlaps, and quite similar shapes, between the three
classifications in the distributions of all parameters.

Figure~\ref{fig:balregcomp2} shows a comparison of the distribution
of Eddington ratios, black hole masses, and bolometric luminosities
for the three samples of quasars (BALs, AALs, and Unabsorbed); mean
values for those quantities are reported in Table~\ref{tab:means}.
While the distributions imply that BAL quasars have systematically
larger bolometric luminosities than unabsorbed quasars, the
distribution of the Eddington ratios for those two samples are quite
similar extending down to very small values
($L_\mathrm{bol}/L_\mathrm{Edd}\lesssim0.1$). This is a rather
profound result given the expectation from the analysis of
Palomar-Green quasars that BALs should have higher than normal
accretion rates, since BAL quasars in the PG sample lie at one
extreme of the \citet{bg92} Eigenvector 1. In the conventional
interpretation, Eigenvector 1 is taken to be driven by the Eddington
ratio \citep[e.g.,][]{boroson02}.

One factor that could affect the robustness of this result is the
average bolometric correction of BAL quasars from 3000\,\AA. If
indeed the spectral shape of BAL quasars is different from
unabsorbed quasars, one can question the validity of using the same
(or similar) bolometric correction. If the mean bolometric
correction for BAL quasars are, for example, higher by factor of
$\sim2$, then the mean Eddington ratio would be of order unity.
However, even if this were the case, we would still have a
significant number of BAL quasars with relatively small Eddington
ratios, given the broad distribution in Fig.~\ref{fig:balregcomp2}.

Regardless, at best it is not clear if the average bolometric
correction for BAL quasars should be larger than that of unabsorbed
quasars. If the difference between BALs and non-BALs is inclination
\citep[see, for example,][for a rebuttal of simple orientation
schemes]{brotherton06}, then one might expect differences since how
one views an axisymmetric object changes the spectrum one observes
\citep[e.g.,][]{kv98}. Optical spectra of BAL do tend to be flatter
than that of unabsorbed quasars \citep[e.g.,][see also
Figure~\ref{fig:balregcomp}]{yv99,bro01,tolea02,reichard03}. On the
other hand, one might question the validity of the black hole mass
scaling relations for BAL quasars, since they have been largely
absent from reverberation mapping campaigns.\footnote{To date, only
one BAL quasar, PG\,1700+518, has had a reported black hole mass
from reverberation mapping \citep{vp06,kaspi05,peterson04,kaspi00}.}
Inaccurate black hole masses would, of course, affect our
computations of the Eddington luminosities.

\begin{figure}
\epsscale{1.0} \plotone{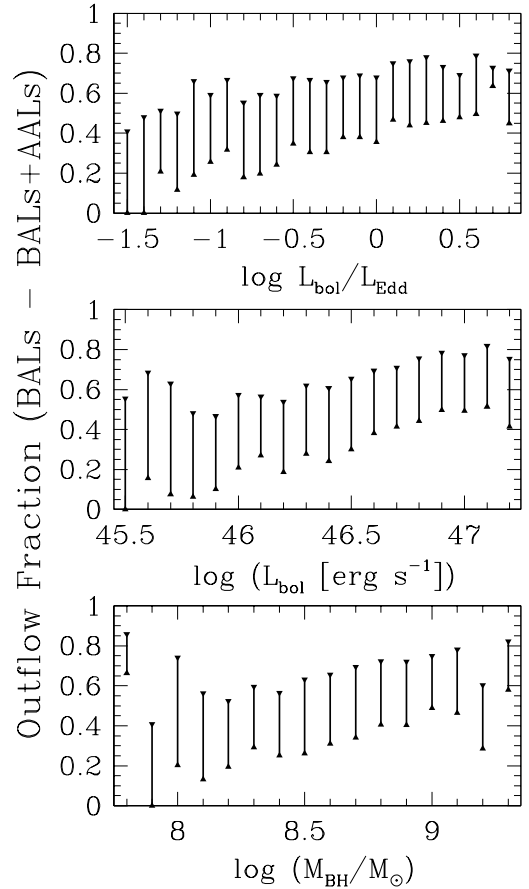} \caption[Outflow Fraction for
Derived Parameters]{We show the fraction of quasars (relative to all
quasars) with high-ionization outflows detected in \ion{C}{4}
absorption as a function of Eddington ratio (top), bolometric
luminosity (middle), and black hole mass (bottom). Lower limits
arise from assuming that only the quasars in the mini-BAL/BAL sample
are BALs, Upper limits arise from assuming that both mini-BAL/BAL
and AAL quasar samples should be counted as outflows.}
\label{fig:balfrac}
\end{figure}

In addition to the comparison in the mean values of the
distributions, we also consider the BAL fraction as a function of
various physical parameters. Figure~\ref{fig:balfrac} shows how the
outflow fraction changes as a function of Eddington ratio,
bolometric luminosity, and black hole mass. In each of the panels,
we provide an upper limit and lower limit to the outflow fraction.
The lower limit assumes that only quasars in our BAL sample should
be counted as outflows:
\begin{equation}
\mathrm{Outflow~Frac. \geq {{BALs} \over {BALs+AALs+Unabsorbed}}}.
\end{equation}
However, it is still possible with our subjective classification
scheme that some outflows (detected in absorption) were placed in
the AAL class. Thus we place an upper limit on the fraction of
quasars with detected outflows by assuming that all objects
classified as BAL or AAL should be counted:
\begin{equation}
\mathrm{Outflow~Frac. \leq {{BALs+AALs} \over
{BALs+AALs+Unabsorbed}}}.
\end{equation}
There are a number of possible origins for AALs (other than
outflowing gas) so the incompleteness of the BAL sample is unlikely
to be very large. Thus, the upper limit should be treated as a very
conservative value.

Generally, there does seem to be a slight trend of increasing
outflow fraction with increasing Eddington ratio, bolometric
luminosity, and black hole mass. However, the distributions are, to
within our conservative upper limits, consistent with being almost
constant.

~\\

\begin{deluxetable*}{llccc}
\tablewidth{0pc}
%
\tablecaption{Sample Mean for Fitted and Derived Parameters}
\tablehead {
\colhead{Quantity} & \colhead {Unit} & \colhead{mini-BALs/BALs} & \colhead{AALs} & \colhead{Unabsorbed}
}
\startdata
$\langle \lambda L_\lambda$(3000\,\AA)$\rangle$ & $10^{45}$\ erg s$^{-1}$          & 7.04$\pm$0.24 (5.67) & 5.88$\pm$0.14 (5.93) & 5.30$\pm$0.11 (5.48) \\
$\langle \alpha_{\mathrm{NUV}} \rangle$         & $F_\lambda\sim\lambda^{-\alpha}$ & 1.29$\pm$0.02 (0.45) & 1.50$\pm$0.01 (0.48) & 1.55$\pm$0.01 (0.42) \\
$\langle$\ion{Mg}{2} FWHM$\rangle$              & $10^3$\ \kms                     & 4.94$\pm$0.06 (1.48) & 4.84$\pm$0.03 (1.33) & 4.75$\pm$0.03 (1.28) \\ \hline \\[-7pt]
$\langle L_\mathrm{bol}/L_\mathrm{Edd} \rangle$ &                                  & 0.46$\pm$0.01 (0.33) & 0.41$\pm$0.01 (0.31) & 0.41$\pm$0.01 (0.32) \\
$\langle M_\mathrm{BH} \rangle$                 & $10^8$\ M$_\odot$                & 6.18$\pm$0.21 (4.95) & 5.16$\pm$0.09 (3.80) & 4.75$\pm$0.07 (3.32) \\
$\langle L_\mathrm{bol} \rangle$                & $10^{46}$\ erg s$^{-1}$          & 3.52$\pm$0.12 (2.84) & 2.94$\pm$0.07 (2.96) & 2.65$\pm$0.05 (2.74)
\enddata
\tablecomments{The mean values for each fitted property, continuum
luminosity, spectra index, and \ion{Mg}{2} $\lambda2800$\ emission-line FWHM
is reported for each quasar class. Parenthetical numbers indicate the
standard deviations of the distributions.}
\label{tab:means}
\end{deluxetable*}

\begin{deluxetable*}{lr@{\,$/$\,}lr@{\,$/$\,}lr@{\,$/$\,}lr@{\,$/$\,}l}
\tablecaption{Spearman Rank Correlation Tests}
\tablehead {
                                    & \multicolumn{2}{c}{BI}    & \multicolumn{2}{c}{AI}                             & \multicolumn{2}{c}{$\vmax$}                  & \multicolumn{2}{c}{$\langle v \rangle$}
}
\startdata
$\lambda L_\lambda$(3000\,\AA)      &  0.047 & 0.286            &              -0.018 & 0.684                        &   {\bf 0.264} & $\mathbf{7.5\times10^{-10}}$ &  {\bf 0.200} & $\mathbf{3.7\times10^{-6}}$   \\
$\alpha_{\mathrm{NUV}}$             &  0.142 & 0.001            &               0.001 & 0.974                        &   {\bf 0.227} & $\mathbf{1.4\times10^{-7}}$  &  {\bf 0.295} & $\mathbf{4.8\times10^{-12}}$  \\
\ion{Mg}{2} FWHM                    & -0.113 & 0.010            &              -0.069 & 0.113                        &  {\bf -0.207} & $\mathbf{1.6\times10^{-6}}$  & {\bf -0.194} & $\mathbf{7.5\times10^{-6}}$   \\ \hline \\[-7pt]
$M_{\mathrm{BH}}$                   & -0.080 & 0.068            &              -0.079 & 0.069                        &       -0.042  & 0.331                        &      -0.063  &          0.150                \\
$L_{\mathrm{bol}}/L_{\mathrm{Edd}}$ &  0.111 & 0.010            &               0.044 & 0.313                        &   {\bf 0.305} & $\mathbf{8.0\times10^{-13}}$ &  {\bf 0.263} & $\mathbf{8.6\times10^{-10}}$  \\ \hline \\[-7pt]
                                    & \multicolumn{2}{c}{Width} & \multicolumn{2}{c}{Max. Depth}                     & \multicolumn{2}{c}{Num. Trough} \\ \hline \\[-7pt]
$\lambda L_\lambda$(3000\,\AA)      &  0.021 & 0.637            &        {\bf -0.282} & $\mathbf{4.8\times10^{-11}}$ &      0.142  & 0.001 \\
$\alpha_{\mathrm{NUV}}$             &  0.058 & 0.186            &             -0.125  & 0.004                        &      0.004  & 0.920 \\
\ion{Mg}{2} FWHM                    & -0.055 & 0.208            & $1.7\times10^{-4}$  & 1.000                        &     -0.103  & 0.018 \\ \hline \\[-7pt]
$M_{\mathrm{BH}}$                   & -0.049 & 0.258            &             -0.160  & $2.3\times10^{-4}$           &     -0.010  & 0.817 \\
$L_{\mathrm{bol}}/L_{\mathrm{Edd}}$ &  0.051 & 0.241             &            -0.148  &          $6.0\times10^{-4}$  &      0.140  & 0.001

\enddata
\tablecomments{For each entry, we list the Spearman rank correlation statistic followed
by the probability of the null hypothesis. The statistics for the bolometric luminosity
are the same as those for $\lambda L_\lambda$(3000\,\AA).}
\label{tab:balspear}
\end{deluxetable*}

\subsection{Correlations With BAL Properties}

In Table~\ref{tab:balspear} we report Spearman rank correlation
statistics (and the corresponding probabilities of the null
hypothesis) for correlations between the measured BAL properties
from \citet[][BI, AI, $\vmax$, $\langle v \rangle$, etc.]{trump06}
and our measured/derived quasar properties [$L_\lambda$(3000\,\AA),
$\alpha_\mathrm{NUV}$, $L_\mathrm{bol}/L_\mathrm{Edd}$, etc.]. In
this correlation table, we have excluded the 11 BAL quasars that do
not appear in the \citet{trump06} catalog. The BALnicity and
absorption indices do not appear to be correlated with any of our
measured/derived quasar properties. This is not surprising given
these quantities, which are essentially variants of an equivalent
width, are complicated functions that depend on the column density
of the outflow, the covering factor of the flow, dilution by
scattered light, and the overall kinematics of the flow.

The single quantity that appears to be correlated with most quasar
properties is the maximum velocity of absorption. Highly significant
(at least in terms of the Spearman rank statistics) correlations are
found between $\vmax$\ and the Eddington ratio and the 3000\,\AA\
luminosity. To a lesser extent, $\vmax$\ also appears to be
correlated with the NUV spectral index. Extreme values of $\vmax$\
are purported to measure the terminal velocity of the mass outflow.
In the current paradigm of radiatively-driven outflows
\citep[e.g.,][]{alb94,mur95,lb02}, the significance of these
correlations is not surprising. Naively, the number of photons
(i.e., luminosity), distribution of momenta (i.e., the spectral
shape), and black hole mass should play a role in determining the
ability to drive an outflow. (Ionization of the gas will also play
role, and we return to this issue later.) Oddly enough, there is not
a significant correlation detected with black hole mass, but there
is with Eddington ratio which incorporates the black hole mass.

\begin{figure}
\epsscale{1.0}
\rotatebox{-90}{\plotone{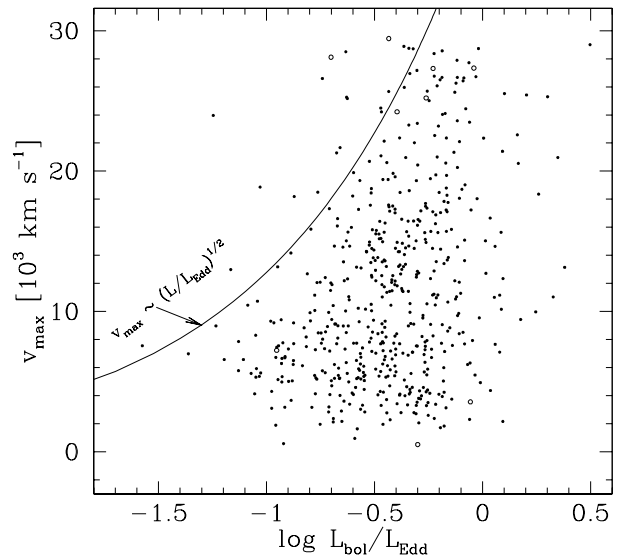}}
\caption[Velocity Correlations]{We show the correlation between the
maximum velocity of absorption and the Eddington ratio. BAL quasars
in the \citet{trump06} catalog are shown as filled circles. New BAL
quasars from this work are shown as unfilled circles. The curve is a
fiducial to show the predicted scaling relationship from
\citet{ham98} and \citet{misawa07}. The normalization of the curve
is arbitrary.} \label{fig:erat}
\end{figure}

\begin{figure*} \epsscale{1.0} \plotone{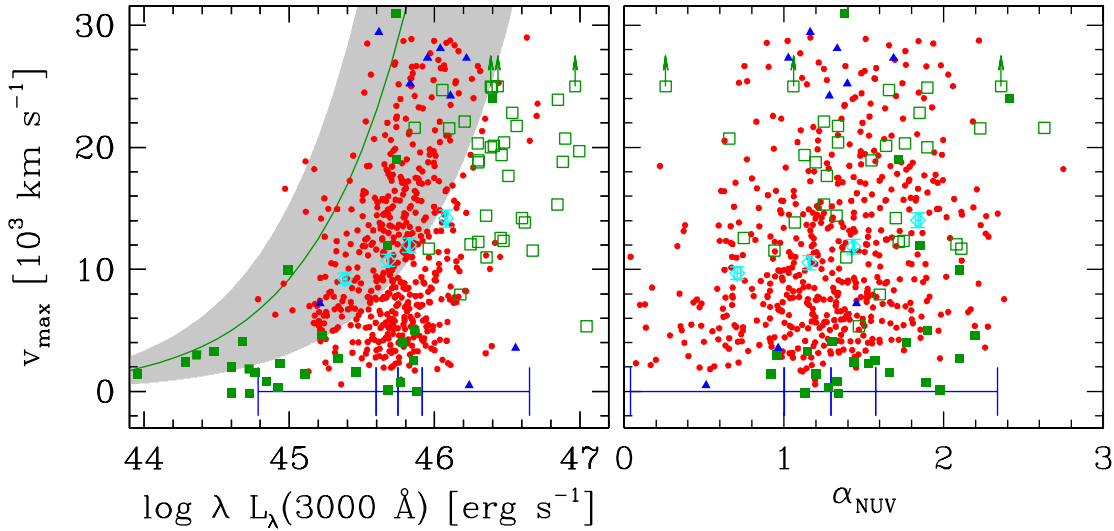}
\caption[Velocity Correlations]{We show the correlation between the
maximum velocity of absorption and 3000\,\AA\ luminosity (left) and
near ultraviolet spectral index (right). BAL quasars in the
\citet{trump06} catalog are shown as (red) circles. New BAL quasars
from this work are shown as (blue) triangles. Filled (green) squares
are the Palomar-Green quasars from \citet{lb02}, while unfilled
(green) squares are LBQS BAL quasars from \citet{gallsc06}. Unfilled
(cyan) diamonds indicate averages of $\vmax$\ and
$L_\lambda$(3000\,\AA)/$\alpha_\mathrm{NUV}$\ for the
\citet{trump06} BALs in bins that are indicated by the ticks on the
(blue) horizontal line. The error bars on the points indicate the
statistical uncertainties in the mean values. The bins were chosen
to have equal numbers of objects (136). The green curve is the
$\vmax$-$L_\lambda$(3000\,\AA) fit from \citet{lb02}, while the
shaded region indicates the reported $1\sigma$\ uncertainty.}
\label{fig:vmax}
\end{figure*}

\subsubsection{$\vmax$\ vs. Eddington Ratio}

\citet{misawa07} and \citet{ham98} show that the terminal velocity
of the outflow should scale with the Eddington ratio
[$v_{\mathrm{terminal}} \sim
(L_\mathrm{bol}/L_\mathrm{Edd})^{1/2}$]. In Figure~\ref{fig:erat},
we plot $\vmax$\ against the our derived Eddington ratio for our BAL
quasars sample (BAL quasars in the \citet{trump06} sample are shown
as filled symbols, while the additional 11 BAL quasars are shown
with unfilled symbols). We show a fiducial curve with the predicted
scaling and an arbitrary normalization. Overall, there does appear
to be an upper envelope that seems to scale as predicted. An actual
empirical fit to the envelope would probably favor a steeper (i.e.,
larger exponent) scaling.

We note one BAL quasar that appears largely discrepant with the
apparent envelope, SDSS J$145408.25+045053.54$\
($z_{\mathrm{em}}=1.98127$). We derived an Eddington ratio of $\log
L_\mathrm{bol}/L_\mathrm{Edd} \approx -1.25$. In addition to the
clear broad \ion{C}{4} absorption-line, \citet{trump06} detect a
narrow \ion{C}{4} absorption-line system ($v_\mathrm{FWHM} \lesssim
500$\,\kms) at $z_\mathrm{abs} = 1.7654$, and report a maximum
velocity of absorption  $\vmax = 23184$\,\kms. However, in addition
to being narrow, this system also shows absorption in a plethora of
low-ionization species (\ion{Al}{2-III}, \ion{Mg}{2}, \ion{Si}{2},
\ion{Si}{4}, \ion{Fe}{2}) and, arguably, should not be counted as
intrinsic absorption. If this system is discounted, then $\vmax$\
for this BAL quasar would be closer to 14000\,\kms, which is more
consistent with the apparent envelope of the other BAL quasars.

\subsubsection{$\vmax$\ vs. Luminosity and Spectral Index}

In addition to the Eddington ratio, the maximum velocity of
absorption also appears to be correlated with the 3000\,\AA\
luminosity and the NUV spectral index. Figure~\ref{fig:vmax} shows
plots of {$\vmax$} against these latter two quantities. In both
panels we distinguish between SDSS BAL quasars that appear in the
\citet{trump06} catalog (plotted as red filled circles) and the 11
new BAL quasars that we added from our subjective search (plotted as
blue filled triangles). In addition, we have plotted the BAL quasars
from the Bright Quasar Survey as tabulated by \citet[][green filled
squares]{lb02}, and from the Large Bright Quasar Survey as tabulated
by \citet[][green open squares.]{gallsc06}.

In the $\vmax-L_\lambda$(3000\,\AA) panel, we also reproduce the fit
(corrected for our cosmology) to the soft X-ray weak PG quasars
carried out by \citet{lb02}. For the updated fit, we have used the
same quasars as \citet[][]{lb02}. We note two differences in our
approach from the \citet{lb02} approach. First, \citet{lb02} fit all
soft X-ray weak quasars in the PG catalog. Soft X-ray weak quasars
are defined as those having $\alpha_\mathrm{ox} \leq -2$, where
$\alpha_\mathrm{ox}$\ is the spectral index derived from the flux at
2500\,\AA\ and 2\,keV. BALs are a subset of soft X-ray weak quasars.
\citet{lb02} noted that soft X-ray weak quasars appeared to define
an upper envelope to the $\vmax-L_\lambda$(3000\,\AA) plot,
consistent with the idea that the absorption does not always trace
the terminal velocity of the outflow. We note here (and discuss
further in \S\ref{sec:discussion}) that, with the inclusion of the
SDSS BALs, it is clear that there is indeed an envelope (though BALs
by themselves do not define it), and our fit should reflect that.
Consequently, we have excluded PG\,2112+059 from the fit since it
clearly does not trace the envelope of the plot (see
Figure~\ref{fig:vmax}, left panel where it is plotted at $\lambda
L_\lambda$(3000\,\AA)$\sim 10^{46.2}$\,erg~s$^{-1}$\ and
$\vmax\sim$24000\,\kms). There a number of BALs at smaller
luminosities that have larger $\vmax$\ values. Furthermore, there
are LBQS BALs at comparable luminosities that show larger $\vmax$\
values. In carrying out the new fit {\it of the envelope}, we have
assumed a 10\% uncertainty in both velocity and luminosity. The
parameters of the revised fit are:
\begin{eqnarray}
v = v_\mathrm{o} (L/L_\mathrm{o})^\alpha, \\
\log L_\mathrm{o}(\mathrm{erg~s^{-1}}) = 45.0, \nonumber \\
\alpha = 0.662 \pm 0.004, \nonumber \\
\log v_\mathrm{o}(\mathrm{\kms}) = 3.96 \pm 0.29. \nonumber
\end{eqnarray}
We have chosen to exclude the SDSS BAL quasars from the fit because
we do not have a convenient criterion for selecting which quasars
define the upper envelope. \citet{lb02} fit all soft X-ray weak
quasars (defined at those having $\alpha_\mathrm{ox} \leq -2$),
since these appeared to define their envelope. This criterion would
not work here for a few reasons:
\begin{itemize}
\item[1.] Since our quasars lie in the redshift range 1.7--2.0,
most of our quasars are not detected in, for example, the {\it
ROSAT} All-Sky Survey. So we do not have a convenient way of
computing $\alpha_\mathrm{ox}$\ {\it en masse} for all of our BAL
quasars.
\item[2.] \citet{strateva05} and \citet{steffen06} have quantified the
dependence of $\alpha_\mathrm{ox}$\ on luminosity
[$\alpha_{\mathrm{ox}} = -0.136 \log L_\nu(2500\,\mathrm{\AA}) +
2.630$]. Consequently, it is possible that soft X-ray weakness as
defined by $\alpha_\mathrm{ox}$\ is not a reasonable criterion for
defining the envelope. From the equation, the range in our
luminosities ($\sim$2\,dex) implies a scatter in the intrinsic
(i.e., unabsorbed) value of $\alpha_\mathrm{ox}$\ of
$\sigma_{\alpha_\mathrm{ox}} \sim 0.27$.
\item[3.] Since all of our objects are pre-selected to have BAL
troughs (even though the maximum observed velocities may be small),
is it possible that all objects may be absorbed in the soft X-ray,
and therefore would be included in the fit if a
$\alpha_\mathrm{ox}$\ criterion were used. This depends on the
detailed relationship between soft X-ray absorption and UV BALs.
See, for example, \citet{gallsc06} for a discussion of this.
\end{itemize}

There is a lot of scatter in each of the panels. So, to bring out
the highly significant correlation, we have taken averages of all
three quantities in four bins. The bins are chosen to have the same
number of points (136) and so they are equally statistically
significant and independent. The error bars indicate the statistical
uncertainties in the mean values.

For the most part, if we interpret our revised
$\vmax-L_\lambda$(3000\,\AA) fit to the \citet{lb02} data as an
upper envelope, the SDSS BAL quasars are very consistent with it,
and fill in the entire range of velocities over the sampled
luminosity range (which overlaps both the BQS and LBQS samples). Of
the 536 BALs, only 2 BAL lies more than $1\sigma$\ above the curve,
and 11 BALs are more deviant than PG\,1700+518, or PG\,1001+054
which were used in reproducing the \citet{lb02} fit.

Another BAL parameter that appears to be correlated with one of the
quasar physical parameters is the maximum depth of the profile (with
both monochromatic and bolometric luminosities). We will examine the
implications of this and correlations between BAL properties (which
is important for interpreting this apparent correlation) in a future
work.

\section{Discussion}
\label{sec:discussion}
\subsection{Are BALs Super-Accretors?}

From the comparison of Eddington ratios between BALs and unabsorbed
quasars, we are led to conclude that BAL quasars are not
super-accretors. While some fraction of BALs appear to accrete near
or above the Eddington limit (51/536 with
$L_\mathrm{bol}/L_\mathrm{Edd} \geq 0.8$), there are a significant
number of BAL quasars that accrete at markedly sub-Eddington rates
(358/536 with $L_\mathrm{bol}/L_\mathrm{Edd} \leq 0.5$).
$L_\mathrm{bol}/L_\mathrm{Edd} \sim 0.5$\ may not be considered a
particularly ``low'' rate, especially in comparison with
lower-luminosity AGN. However, it is important to note that BALs do
not segregate themselves from other AGN in their distribution of
Eddington ratios (Fig.~\ref{fig:balregcomp2}) and the fraction of
quasars with BALs remains significant down to ``low'' Eddington
ratios ($\sim 0.03$, Fig.\ref{fig:balfrac}).

While Fig.~\ref{fig:balfrac} apparently shows a BAL fraction of
$\sim$0 at lower Eddington ratios, this is only an artifact of small
number statistics. We only have six quasars in our sample with
$L_\mathrm{bol}/L_\mathrm{Edd} \leq 0.03$, and one is a BAL, giving
a BAL fraction of 16.7\%. Moreover, very low luminosity quasars may
not be capable of driving an outflow with a large velocity
dispersion, but may still have significant outflows. We note that
four of the six $L_\mathrm{bol}/L_\mathrm{Edd} \leq 0.03$\ quasars
do show AALs.

This is somewhat puzzling given that supposedly BAL quasars lie at
one extreme of the \citet{bg92} Eigenvectors 1 and 2
\citep{bg92,yw03}. \citet{boroson02} interpreted the Eddington ratio
and the absolute mass accretion rate as the principal drivers of
these two eigenvectors (see their Figure~7 which provides an
interpretive diagram). BAL quasars are thought to occupy the high
$L_\mathrm{bol}/L_\mathrm{Edd}$-high $\dot{M_\mathrm{acc}}$\ region
of this diagram. Our result apparently contradicts this idea.

There are two issues here in understanding why the \citet{boroson02}
interpretation may be too simple: (1) the BALs in the Boroson sample
may not sample the full range of Eigenvector 1 properties allowed by
the BAL parent population, and (2) the interpretation of Eddington
ratio as the principal driver of Eigenvector 1 may not be correct
(or complete).

There are only $\sim4$\ BALs in the sample of 162 objects employed
in the Boroson analysis. Thus, it is possible that these four
objects simply do not sample the full range of Eigenvectors 1 and 2
that is allowed by the parent population of BAL quasars. (In
comparison, we have increased the sample by more than a couple
orders of magnitude.) On the other hand, from an analysis of 11
$z\sim2$\ BAL quasars, \citet{yw03} find that they do seem to lie on
one extreme of the \ion{Fe}{2} -- [\ion{O}{3}] distribution. Since
this correlation is one of primary constituents of Eigenvector 1,
they conclude that $z\sim2$\ BAL quasars lie at one extreme of
Eigenvector 1 like their low-z kin. We note, however, that another
ingredient of Eigenvector 1 is the H$\beta$\ FWHM. From Figure~2 of
\citet{yw03}, it appears that the range of H$\beta$\ FWHM for BAL
quasars overlaps completely with that of non-BAL quasars, so this
assertion is not strictly true. BAL quasars at higher redshift are
not simply extreme Eigenvector 1 objects.

We note that all objects in the \citet{boroson02} sample lie at low
redshift ($z \lesssim 0.8$), while all of our objects lie at higher
redshift ($1.7 \lesssim z \lesssim 2$). It is possible that there is
some form of evolution in the population of BAL quasars (e.g.,
luminosity changes or differences in black hole mass) such that low
redshift objects have systematically high Eddington ratios while
higher redshift BAL quasars do not. It is also possible the first
principal component of quasar optical properties changes with
redshift. Our result is further supported by \citet{yw03}, who find
$z\sim2$\ BAL quasars with Eddington ratios as low as $\sim0.17$.
This implies that accretion rate is not the principal driver of
Eigenvector 1 properties (or that it is a secondary effect that is
true only at low-z).

\citet{yw03} speculate that the principal driver of Eigenvector 1
may instead be the availability of ``cold'' gas fueling the
accretion disk as implied, perhaps, by the strong \ion{Fe}{2} and
weaker [\ion{O}{3}] emission. In this interpretation, the high
correlation between Eigenvector 1 and the Eddington ratio in low
redshift PG sample is seen as a secondary correlation. At low
redshift, most objects have a low fuel supply, hence low Eddington
ratio, and the few that have an abundance of cold fuel are able to
accrete near the Eddington rate. These low redshift, high Eddington
ratio black holes, are those that can exhibit broad absorption
lines. At high redshift, however, most quasars (i.e., both BAL and
non-BAL quasars) have an ample supply of cold gas with BAL quasars
having the largest supply (perhaps because they are the youngest
sources). Consequently, all high redshift quasars are able to
accrete near their Eddington limits, hence having similar (and
``high'') Eddington ratios. Those that have the largest supply of
cold gas, and hence largest fueling rates, have the largest outflow
rate and exhibit broad absorption lines.

We note that an important component of the \citet{yw03}
interpretation is the differences in black hole mass between their
high redshift sample and the low redshift PG sample. The high
redshift quasars from \cite{yw03} have masses $\gtrsim
10^9$\,M$_\odot$, whereas the low redshift PG quasars have lower
masses ($<10^9$\,M$_\odot$). It is this fact that allows the the
high redshift quasars to accommodate higher absolute accretion rates
(since they have larger Eddington accretion rates) and have higher
luminosities than the low redshift PG quasars.

Our sample of quasars from SDSS is at similar redshifts to those in
the \citet{yw03} sample ($z\sim2$), but have smaller black hole
masses, smaller bolometric luminosities and span a wider range of
Eddington ratios. Compared to the low redshift PG sample, however,
we have larger black hole masses, larger bolometric luminosities,
but comparable range of Eddington ratios. Since we do not have
rest-frame optical spectra for these quasar, their Eigenvector 1
properties are not known directly. Further study (e.g., infrared
spectroscopy) is needed to understand how the Eigenvector 1
properties of these objects relate to the low redshift PG sample and
the high redshift \citet{yw03} sample.

Nevertheless, if we take at face value the fact that we see BAL
quasars in 16.7\% (and outflows in as many as 83\%) of
$L_\mathrm{bol}/L_\mathrm{Edd} \leq 0.03$\ objects, then we must
question the completeness of the \citet{yw03} interpretation. High
redshift BAL and non-BAL quasars do have similar Eddington ratios,
but not because both populations are accreting near their Eddington
limits. Even if we accept that BAL quasars do indeed lie at one
extreme of the [\ion{O}{3}] $\lambda$5007 equivalent width -
\ion{Fe}{2}/H$\beta$\ intensity ratio anti-correlation, we must
understand first the detailed physics governing that
anti-correlation and drivers behind it. \citet{yw03} have shown that
it is not simply the Eddington ratio. The addition of this work
questions if it is simply the fueling rate. There are other effects,
such as the covering factor of the outflow and underlying cause(s)
of that covering factor, that must be considered in this
interpretation.

\subsection{Radiatively-driven winds?}

Our second result is a confirmation of the luminosity-dependent
envelope to the maximum velocity of absorption \citep{lb02}. From
Figure~\ref{fig:vmax}, the maximum velocities of our BAL quasars
appear to obey the $v \sim L^{0.66}$\ best-fit relation re-derived
with the corrected cosmology. To explain the slope, those authors
consider two formulations of terminal velocity and its dependence on
luminosity: $\vmax \propto \sqrt{\Gamma L/R}$, where $L$\ is the
luminosity, $R$\ is the launching radius of the wind, and $\Gamma$\
is the force-multiplier \citep{cak75}. Aside from the explicit
$\sqrt{L}$\ factor, the luminosity also comes into play in
determining both the force-multiplier and the launching radius. One
can assume that the launching radius is independent of luminosity or
that there is a scaling (e.g., $R \sim \sqrt{L}$). This gives a
range of velocity scalings: $\vmax \sim \sqrt{\Gamma} L^{0.25-0.5}$.
The original \citet{lb02} best fit had a steeper slope which led the
authors to conclude that there was a luminosity dependence to the
force-multiplier. We agree with that conclusion with a
cosmology-corrected revision of the required dependence (combining
the theoretical dependence of $\vmax$\ on $\Gamma$\ and luminosity
with our revised best-fit to the $\vmax-L_\lambda$(3000\,\AA)
envelope): $\Gamma \sim L^{0.32-0.82}$.

The force multiplier, $\Gamma$, relates the force from a line-driven
(or edge-driven) wind to one that is driven purely by electron
(Thompson) scattering. UV transitions (e.g., in the range
200-3200\,\AA) are thought to be principal contributors to
line-driving. The precise relationship between the force multiplier
and the luminosity (or, alternatively, ionization parameter) can be
complicated \citep[e.g., ][]{alb94}, and is beyond the scope of this
paper. It is sufficient here to say that a relationship between the
force-multiplier and the luminosity is expected (i.e., the two
quantities are not {\it a priori} expected to be independent).

\citet{sulentic06} employ the semi-empirical $\Gamma(U)$\
relationship derived by \citet[][$\log \Gamma \approx 2.551 - 0.536
\log U$]{al94} to explicitly characterize the dependence of $\vmax$\
with ionization parameter ($\vmax \sim U^{-1/4}$). We clarify here
that this relationship is only valid under certain conditions which
are not necessarily valid throughout a BAL flow: (1) the
relationship does not account for optical depth effects\footnote{The
full relation given in \citet{al94} does include terms accounting
for optical depth, but these were not included in the
\citet{sulentic06} analysis.} (all lines are assumed to be highly
optically thin), (2) the force multiplier only takes into account
line-driving (driving from edges may also be important), (3) the
driving spectrum is that derived by \citet{mf87} which may not be
generally applicable, and (4) the range of valid ionization
parameters is small ($-1.5 \leq \log U \leq +0.5$). A revision of
this relation given in \citet*{alb94} expanded the range of
ionization parameter to $-3 \leq \log U \leq 1$, but with a more
complicated semi-empirical fit: $\log \Gamma = 3.642 - 0.1445 [\log
U + 3]^2$. [Note that this still uses the \citet{mf87} spectrum and
also does not account for optical depth effects.] Furthermore,
\citet{psd98} and \citet{pk04} note that the force-multiplier is
very sensitive to both the Eddington ratio and to the black hole
mass. A line-driven wind cannot be formed unless
$L_\mathrm{bol}/L_\mathrm{Edd} \gtrsim \Gamma^{-1}$.

Another complication in the evaluation of the force-multiplier (as
we have noted) and its effect on the terminal velocity of the wind
is the dependence on the shape of the driving spectrum. From our
fits, we have a hint of this importance in the apparent correlation
between $\vmax$\ and $\alpha_\mathrm{NUV}$. The correlation
(Fig.~\ref{fig:vmax}, right panel) is weak, but statistically
significant (Table~\ref{tab:balspear}). The only resonant lines in
the NUV region of the spectrum are
\ion{Mg}{2}$\lambda\lambda$2796,2803 and the plethora of \ion{Fe}{2}
lines in the range 2230--2600\,\AA. In the high-ionization BALs in
this analysis, it is unlikely the these transitions play an
important role in line-driving. A more likely cause for the
correlation is the loose dependence between $\alpha_\mathrm{NUV}$\
and the UV flux at shorter wavelengths, in the sense that ``harder''
values of $\alpha_\mathrm{NUV}$\ will yield relatively more UV/FUV
photons than ``softer'' values. [Note that, with our sign
convention, $F_\lambda \sim \lambda^{-\alpha}$, a larger value of
$\alpha_\mathrm{NUV}$\ implies a harder spectrum.] We note here
that, in our sample, $\alpha_\mathrm{NUV}$\ is anti-correlated with
$L_\lambda$(3000\,\AA) (i.e., more luminous sources are ``softer'').
Thus the correlation between $\vmax$\ and $L_\lambda$(3000\,\AA)
contributes to the weakness in the correlation between $\vmax$\ and
$\alpha_\mathrm{NUV}$. Consequently, the dependence of $\vmax$\ on
$\alpha_\mathrm{NUV}$\ is probably stronger than indicated.

Finally, we note that $\vmax$ - $\lambda L_\lambda$(3000\,\AA) curve
truly constitutes an upper envelope. A significant fraction of
points lie well below the curve. The details in the computation of
the force-multiplier likely can explain some of this. However, there
are a few other possibilities to note. As with all objects that do
not show spherical symmetry, the orientation relative to the
observer's sight-line probably plays a role. The magnitude of this
effect is not clear, and is affected by the fact that we do not yet
have a full theoretical exploration of the accretion disk-wind
paradigm. [For example, the orientation of radio-jets and
polarization of light in the troughs in some radio-loud BALs seems
to indicate a range of accretion-disk orientations
\citep{zhou06,brotherton06}, but the opening angle of the wind may
also be important.] Moreover, while the existence of the $\vmax$ -
$\lambda L_\lambda$(3000\,\AA) envelope indicates the importance of
radiation-pressure in driving the BAL outflow, this is not the only
possible mechanism. Magnetically-driven \citep[e.g.,][]{everett05},
and thermally-driven \citep[e.g.,][]{kk01} winds may explain some
fraction of BALs with smaller observed values of $\vmax$.

For the \cite{lb02} points that appear at low $\vmax$, one must also
question the location of the absorbing gas. It is possible that some
of these may be unrelated to the immediate quasar environment, and
may arise from the ISM of the host galaxy, or perhaps other nearby
galaxies. However, this cannot explain the objects from our sample
that have small $\vmax$. As we noted before, all of our objects are
selected because they show broad wind-like profiles. Absorption from
the host galaxy or nearby structures would have velocity dispersions
of a few hundred $\kms$\ appearing close to the quasar redshift, and
thus would have been relegated to our AAL class. The low-velocity
objects in our sample may indicate the importance of non-radiative
processes like thermal gradients \citep[e.g.,][]{kk01} or
magneto-centrifugal rotation \citep[e.g.,][]{everett05} in driving
outflows.

A further complication is the apparently redshifted absorption that
we have in some of our BALs. In the case of host galaxy ISM
absorption, this would not be a problem since the \ion{C}{4}
$\lambda$1549 emission line is known to be blueshifted with respect
to the systemic redshift
\citep[e.g.,][]{gaskell83,esp93,rich02b,vdb01}. However, in the
accretion-disk/wind paradigm, the \ion{C}{4} broad emission and
broad absorption lines are thought to be produced by different parts
of the same flow \citep{mur95,baldwin96,mur97,elvis00}. [Some
fraction of the \ion{C}{4} $\lambda$1549 emission line may arise
from different regions, like a virialized component
\citep[e.g.,][]{wills93,bro94}, but seems to be dominated by a wind
component.] A full interpretation of how (or whether) this situation
(apparently redshifted broad absorption sitting on top of a
blueshifted broad emission line) is complicated by the uncertainty
in what part or parts of the wind are being sampled by the observed
absorption and in what determines the velocity where the emissivity
peaks. Both of these questions rely on the orientations of the
observer relative to the axis of the disk and to the direction (and
opening angle) of the wind, as well as the dynamics and ionization
structure of the wind.

\section{Summary}
\label{sec:summary}
\begin{itemize}
\item [1.] We have subjectively scrutinized a sample of 5088
$1.7 \leq z \leq 2$\ quasars from the Second Data Release of the
Sloan Digital Sky Survey and placed the objects into three classes:
562 objects show clear signs of an outflow; 2573 show no signs of an
outflow (in absorption); 1898 show evidence of ``associated'' narrow
absorption which could have several different locations. The
frequency of observed outflows is different from the recent AI-based
BAL catalog of \citet{trump06} and we point out several differences.
\item[2.] We have estimated black hole masses and Eddington ratios
for all quasars in this sample. We find that there is no appreciable
difference in either the black hole mass or Eddington ratio
distribution between BALs and non-BALs. This implies that BALs in
this redshift range are not super-accretors. Like \cite{yw03}, we
speculate that Eddington ratio may not be the principal driver of
the \citet{bg92} Eigenvector 1, but rather availability of cold gas,
and the ``down-sizing'' evolution of black-hole mass in accreting
systems.
\item[3.] We find that the maximum velocity of absorption as a
function of luminosity has an upper envelope that is consistent with
the best fit from \citet{lb02}. This upper envelope is easily
interpreted as the terminal velocity of radiatively-driven wind. We
find that it is also correlated with NUV spectral index, which may
indicate the importance of the SED shape in governing the dynamics
of the outflow. However, many of our BALs terminate at small
velocities. This may indicate the importance of wind-orientation, or
non-radiative processes in driving outflows.
\end{itemize}

\acknowledgments

We wish to thank Sarah Gallagher, Toru Misawa, Ari Laor, Jonathan
Trump, Gordon Richards, and John Everett for helpful discussions.
Special thanks are given to the referee Pat Hall for an excellent,
thorough, and well-thought out review of the paper. This work was
supported by the National Science Foundation under Grant No.
0507781.

Funding for the SDSS and SDSS-II has been provided by the Alfred P.
Sloan Foundation, the Participating Institutions, the National
Science Foundation, the U.S. Department of Energy, the National
Aeronautics and Space Administration, the Japanese Monbukagakusho,
the Max Planck Society, and the Higher Education Funding Council for
England. The SDSS Web Site is http://www.sdss.org/.

The SDSS is managed by the Astrophysical Research Consortium for the
Participating Institutions. The Participating Institutions are the
American Museum of Natural History, Astrophysical Institute Potsdam,
University of Basel, University of Cambridge, Case Western Reserve
University, University of Chicago, Drexel University, Fermilab, the
Institute for Advanced Study, the Japan Participation Group, Johns
Hopkins University, the Joint Institute for Nuclear Astrophysics,
the Kavli Institute for Particle Astrophysics and Cosmology, the
Korean Scientist Group, the Chinese Academy of Sciences (LAMOST),
Los Alamos National Laboratory, the Max-Planck-Institute for
Astronomy (MPIA), the Max-Planck-Institute for Astrophysics (MPA),
New Mexico State University, Ohio State University, University of
Pittsburgh, University of Portsmouth, Princeton University, the
United States Naval Observatory, and the University of Washington.


\begin{thebibliography}{}

\bibitem[\protect\citeauthoryear{{Abazajian} et~al.}{{Abazajian}
  et~al.}{2004}]{sdss2}
{Abazajian}, K., et~al. 2004, \aj, 128, 502

\bibitem[\protect\citeauthoryear{{Arav} \& {Li}}{{Arav} \& {Li}}{1994}]{al94}
{Arav}, N.,  \& {Li}, Z.-Y. 1994, \apj, 427, 700

\bibitem[\protect\citeauthoryear{{Arav}, {Li}, \& {Begelman}}{{Arav}
  et~al.}{1994}]{alb94}
{Arav}, N., {Li}, Z.-Y.,  \& {Begelman}, M.~C. 1994, \apj, 432, 62

\bibitem[\protect\citeauthoryear{{Baldwin} et~al.}{{Baldwin}
  et~al.}{1996}]{baldwin96}
{Baldwin}, J.~A., et~al. 1996, \apj, 461, 664

\bibitem[\protect\citeauthoryear{{Becker} et~al.}{{Becker}
  et~al.}{2000}]{becker00}
{Becker}, R.~H., {White}, R.~L., {Gregg}, M.~D., {Brotherton},
M.~S.,
  {Laurent-Muehleisen}, S.~A.,  \& {Arav}, N. 2000, \apj, 538, 72

\bibitem[\protect\citeauthoryear{{Boroson}}{{Boroson}}{2002}]{boroson02}
{Boroson}, T.~A. 2002, \apj, 565, 78

\bibitem[\protect\citeauthoryear{{Boroson} \& {Green}}{{Boroson} \&
  {Green}}{1992}]{bg92}
{Boroson}, T.~A.,  \& {Green}, R.~F. 1992, \apjs, 80, 109

\bibitem[\protect\citeauthoryear{{Brandt} \& {Gallagher}}{{Brandt} \&
  {Gallagher}}{2000}]{bg00}
{Brandt}, W.~N.,  \& {Gallagher}, S.~C. 2000, New Astronomy Review,
44, 461

\bibitem[\protect\citeauthoryear{{Brotherton}, {de Breuck}, \&
  {Schaefer}}{{Brotherton} et~al.}{2006}]{brotherton06}
{Brotherton}, M.~S., {de Breuck}, C.,  \& {Schaefer}, J.~J. 2006,
\mnras, 372,
  L58

\bibitem[\protect\citeauthoryear{{Brotherton} et~al.}{{Brotherton}
  et~al.}{2001}]{bro01}
{Brotherton}, M.~S., {Tran}, H.~D., {Becker}, R.~H., {Gregg}, M.~D.,
  {Laurent-Muehleisen}, S.~A.,  \& {White}, R.~L. 2001, \apj, 546, 775

\bibitem[\protect\citeauthoryear{{Brotherton} et~al.}{{Brotherton}
  et~al.}{1999}]{bro99}
{Brotherton}, M.~S., et~al. 1999, \apjl, 520, L87

\bibitem[\protect\citeauthoryear{{Brotherton} et~al.}{{Brotherton}
  et~al.}{1994}]{bro94}
{Brotherton}, M.~S., {Wills}, B.~J., {Steidel}, C.~C.,  \&
{Sargent}, W.~L.~W.
  1994, \apj, 423, 131

\bibitem[\protect\citeauthoryear{{Castor}, {Abbott}, \& {Klein}}{{Castor}
  et~al.}{1975}]{cak75}
{Castor}, J.~I., {Abbott}, D.~C.,  \& {Klein}, R.~I. 1975, \apj,
195, 157

\bibitem[\protect\citeauthoryear{{Dietrich} \& {Hamann}}{{Dietrich} \&
  {Hamann}}{2004}]{dm04}
{Dietrich}, M.,  \& {Hamann}, F. 2004, \apj, 611, 761

\bibitem[\protect\citeauthoryear{{Elvis}}{{Elvis}}{2000}]{elvis00}
{Elvis}, M. 2000, \apj, 545, 63

\bibitem[\protect\citeauthoryear{{Espey}}{{Espey}}{1993}]{esp93}
{Espey}, B.~R. 1993, \apjl, 411, L59

\bibitem[\protect\citeauthoryear{{Everett}}{{Everett}}{2005}]{everett05}
{Everett}, J.~E. 2005, \apj, 631, 689

\bibitem[\protect\citeauthoryear{{Gallagher} et~al.}{{Gallagher}
  et~al.}{2006}]{gallsc06}
{Gallagher}, S.~C., {Brandt}, W.~N., {Chartas}, G., {Priddey}, R.,
{Garmire},
  G.~P.,  \& {Sambruna}, R.~M. 2006, \apj, 644, 709

\bibitem[\protect\citeauthoryear{{Gallagher} et~al.}{{Gallagher}
  et~al.}{1999}]{gallsc}
{Gallagher}, S.~C., {Brandt}, W.~N., {Sambruna}, R.~M., {Mathur},
S.,  \&
  {Yamasaki}, N. 1999, \apj, 519, 549

\bibitem[\protect\citeauthoryear{{Gaskell}}{{Gaskell}}{1983}]{gaskell83}
{Gaskell}, C.~M. 1983, \apjl, 267, L1

\bibitem[\protect\citeauthoryear{{Goodrich}}{{Goodrich}}{1997}]{goodrich97}
{Goodrich}, R.~W. 1997, \apj, 474, 606

\bibitem[\protect\citeauthoryear{{Goodrich} \& {Miller}}{{Goodrich} \&
  {Miller}}{1995}]{good95}
{Goodrich}, R.~W.,  \& {Miller}, J.~S. 1995, \apjl, 448, L73

\bibitem[\protect\citeauthoryear{{Gregg} et~al.}{{Gregg}
  et~al.}{2000}]{gregg00}
{Gregg}, M.~D., {Becker}, R.~H., {Brotherton}, M.~S.,
{Laurent-Muehleisen},
  S.~A., {Lacy}, M.,  \& {White}, R.~L. 2000, \apj, 544, 142

\bibitem[\protect\citeauthoryear{{Gregg}, {Becker}, \& {de Vries}}{{Gregg}
  et~al.}{2006}]{gbd06}
{Gregg}, M.~D., {Becker}, R.~H.,  \& {de Vries}, W. 2006, \apj, 641,
210

\bibitem[\protect\citeauthoryear{{Hall} et~al.}{{Hall} et~al.}{2002}]{hallai}
{Hall}, P.~B., et~al. 2002, \apjs, 141, 267

\bibitem[\protect\citeauthoryear{{Hamann}}{{Hamann}}{1998}]{ham98}
{Hamann}, F. 1998, \apj, 500, 798

\bibitem[\protect\citeauthoryear{{Kaspi} et~al.}{{Kaspi}
  et~al.}{2005}]{kaspi05}
{Kaspi}, S., {Maoz}, D., {Netzer}, H., {Peterson}, B.~M.,
{Vestergaard}, M.,
  \& {Jannuzi}, B.~T. 2005, \apj, 629, 61

\bibitem[\protect\citeauthoryear{{Kaspi} et~al.}{{Kaspi}
  et~al.}{2000}]{kaspi00}
{Kaspi}, S., {Smith}, P.~S., {Netzer}, H., {Maoz}, D., {Jannuzi},
B.~T.,  \&
  {Giveon}, U. 2000, \apj, 533, 631

\bibitem[\protect\citeauthoryear{{Kriss}}{{Kriss}}{1994}]{specfit}
{Kriss}, G. 1994, in ASP Conf. Ser. 61: Astronomical Data Analysis
Software and
  Systems III, ed. D.~R. {Crabtree}, R.~J. {Hanisch}, \& J.~{Barnes}, 437

\bibitem[\protect\citeauthoryear{{Krolik}}{{Krolik}}{1999}]{krolikagn}
{Krolik}, J.~H. 1999, {Active Galactic Nuclei : From The Central
Black Hole To
  The Galactic Environment} (Princeton, N.~J.~: Princeton University Press,
  c1999.)

\bibitem[\protect\citeauthoryear{{Krolik} \& {Kriss}}{{Krolik} \&
  {Kriss}}{2001}]{kk01}
{Krolik}, J.~H.,  \& {Kriss}, G.~A. 2001, \apj, 561, 684

\bibitem[\protect\citeauthoryear{{Krolik} \& {Voit}}{{Krolik} \&
  {Voit}}{1998}]{kv98}
{Krolik}, J.~H.,  \& {Voit}, G.~M. 1998, \apjl, 497, L5

\bibitem[\protect\citeauthoryear{{Laor} \& {Brandt}}{{Laor} \&
  {Brandt}}{2002}]{lb02}
{Laor}, A.,  \& {Brandt}, W.~N. 2002, \apj, 569, 641

\bibitem[\protect\citeauthoryear{{Mathews} \& {Ferland}}{{Mathews} \&
  {Ferland}}{1987}]{mf87}
{Mathews}, W.~G.,  \& {Ferland}, G.~J. 1987, \apj, 323, 456

\bibitem[\protect\citeauthoryear{{McLure} \& {Jarvis}}{{McLure} \&
  {Jarvis}}{2002}]{mj02}
{McLure}, R.~J.,  \& {Jarvis}, M.~J. 2002, \mnras, 337, 109

\bibitem[\protect\citeauthoryear{{Misawa} et~al.}{{Misawa}
  et~al.}{2007}]{misawa07}
{Misawa}, T., {Charlton}, J.~C., {Eracleous}, M., {Ganguly}, R.,
{Tytler}, D.,
  {Kirkman}, D., {Suzuki}, N.,  \& {Lubin}, D. 2007, \apj, {in press
  (astro-ph/0702101)}

\bibitem[\protect\citeauthoryear{{Murray} \& {Chiang}}{{Murray} \&
  {Chiang}}{1997}]{mur97}
{Murray}, N.,  \& {Chiang}, J. 1997, \apj, 474, 91

\bibitem[\protect\citeauthoryear{{Murray} et~al.}{{Murray}
  et~al.}{1995}]{mur95}
{Murray}, N., {Chiang}, J., {Grossman}, S.~A.,  \& {Voit}, G.~M.
1995, \apj,
  451, 498

\bibitem[\protect\citeauthoryear{{Najita}, {Dey}, \& {Brotherton}}{{Najita}
  et~al.}{2000}]{ndb00}
{Najita}, J., {Dey}, A.,  \& {Brotherton}, M. 2000, \aj, 120, 2859

\bibitem[\protect\citeauthoryear{{Peterson} et~al.}{{Peterson}
  et~al.}{2004}]{peterson04}
{Peterson}, B.~M., et~al. 2004, \apj, 613, 682

\bibitem[\protect\citeauthoryear{{Proga} \& {Kallman}}{{Proga} \&
  {Kallman}}{2004}]{pk04}
{Proga}, D.,  \& {Kallman}, T.~R. 2004, \apj, 616, 688

\bibitem[\protect\citeauthoryear{{Proga}, {Stone}, \& {Drew}}{{Proga}
  et~al.}{1998}]{psd98}
{Proga}, D., {Stone}, J.~M.,  \& {Drew}, J.~E. 1998, \mnras, 295,
595

\bibitem[\protect\citeauthoryear{{Reichard} et~al.}{{Reichard}
  et~al.}{2003a}]{reichard03b}
{Reichard}, T.~A., et~al. 2003a, \aj, 126, 2594

\bibitem[\protect\citeauthoryear{{Reichard} et~al.}{{Reichard}
  et~al.}{2003b}]{reichard03}
{Reichard}, T.~A., et~al. 2003b, \aj, 125, 1711

\bibitem[\protect\citeauthoryear{{Richards} et~al.}{{Richards}
  et~al.}{2003}]{gtr03}
{Richards}, G.~T., et~al. 2003, \aj, 126, 1131

\bibitem[\protect\citeauthoryear{{Richards} et~al.}{{Richards}
  et~al.}{2006}]{gtr06}
{Richards}, G.~T., et~al. 2006, \apjs, 166, 470

\bibitem[\protect\citeauthoryear{{Richards} et~al.}{{Richards}
  et~al.}{2002}]{rich02b}
{Richards}, G.~T., {Vanden Berk}, D.~E., {Reichard}, T.~A., {Hall},
P.~B.,
  {Schneider}, D.~P., {SubbaRao}, M., {Thakar}, A.~R.,  \& {York}, D.~G. 2002,
  \aj, 124, 1

\bibitem[\protect\citeauthoryear{{Sanders} et~al.}{{Sanders}
  et~al.}{1988}]{sanders88}
{Sanders}, D.~B., {Soifer}, B.~T., {Elias}, J.~H., {Neugebauer}, G.,
\&
  {Matthews}, K. 1988, \apjl, 328, L35

\bibitem[\protect\citeauthoryear{{Shang} et~al.}{{Shang}
  et~al.}{2005}]{shang05}
{Shang}, Z., et~al. 2005, \apj, 619, 41

\bibitem[\protect\citeauthoryear{{Sprayberry} \& {Foltz}}{{Sprayberry} \&
  {Foltz}}{1992}]{sf92}
{Sprayberry}, D.,  \& {Foltz}, C.~B. 1992, \apj, 390, 39

\bibitem[\protect\citeauthoryear{{Steffen} et~al.}{{Steffen}
  et~al.}{2006}]{steffen06}
{Steffen}, A.~T., {Strateva}, I., {Brandt}, W.~N., {Alexander},
D.~M.,
  {Koekemoer}, A.~M., {Lehmer}, B.~D., {Schneider}, D.~P.,  \& {Vignali}, C.
  2006, \aj, 131, 2826

\bibitem[\protect\citeauthoryear{{Strateva} et~al.}{{Strateva}
  et~al.}{2005}]{strateva05}
{Strateva}, I.~V., {Brandt}, W.~N., {Schneider}, D.~P., {Vanden
Berk}, D.~G.,
  \& {Vignali}, C. 2005, \aj, 130, 387

\bibitem[\protect\citeauthoryear{{Sulentic} et~al.}{{Sulentic}
  et~al.}{2006}]{sulentic06}
{Sulentic}, J.~W., {Dultzin-Hacyan}, D., {Marziani}, P., {Bongardo},
C.,
  {Braito}, V., {Calvani}, M.,  \& {Zamanov}, R. 2006, Revista Mexicana de
  Astronomia y Astrofisica, 42, 23

\bibitem[\protect\citeauthoryear{{Tolea}, {Krolik}, \& {Tsvetanov}}{{Tolea}
  et~al.}{2002}]{tolea02}
{Tolea}, A., {Krolik}, J.~H.,  \& {Tsvetanov}, Z. 2002, \apjl, 578,
L31

\bibitem[\protect\citeauthoryear{{Trump} et~al.}{{Trump}
  et~al.}{2006}]{trump06}
{Trump}, J.~R., et~al. 2006, \apjs, 165, 1

\bibitem[\protect\citeauthoryear{{Vanden Berk} et~al.}{{Vanden Berk}
  et~al.}{2001}]{vdb01}
{Vanden Berk}, D.~E., et~al. 2001, \aj, 122, 549

\bibitem[\protect\citeauthoryear{{Vestergaard}}{{Vestergaard}}{2004}]{ves04}
{Vestergaard}, M. 2004, \apj, 601, 676

\bibitem[\protect\citeauthoryear{{Vestergaard} \& {Peterson}}{{Vestergaard} \&
  {Peterson}}{2006}]{vp06}
{Vestergaard}, M.,  \& {Peterson}, B.~M. 2006, \apj, 641, 689

\bibitem[\protect\citeauthoryear{{Vestergaard} \& {Wilkes}}{{Vestergaard} \&
  {Wilkes}}{2001}]{vw01}
{Vestergaard}, M.,  \& {Wilkes}, B.~J. 2001, \apjs, 134, 1

\bibitem[\protect\citeauthoryear{{Voit}, {Weymann}, \& {Korista}}{{Voit}
  et~al.}{1993}]{vwk93}
{Voit}, G.~M., {Weymann}, R.~J.,  \& {Korista}, K.~T. 1993, \apj,
413, 95

\bibitem[\protect\citeauthoryear{{Weymann} et~al.}{{Weymann}
  et~al.}{1991}]{weymann91}
{Weymann}, R.~J., {Morris}, S.~L., {Foltz}, C.~B.,  \& {Hewett},
P.~C. 1991,
  \apj, 373, 23

\bibitem[\protect\citeauthoryear{{Wills} et~al.}{{Wills}
  et~al.}{1993}]{wills93}
{Wills}, B.~J., {Brotherton}, M.~S., {Fang}, D., {Steidel}, C.~C.,
\&
  {Sargent}, W.~L.~W. 1993, \apj, 415, 563

\bibitem[\protect\citeauthoryear{{Wise} et~al.}{{Wise} et~al.}{2004}]{wise04}
{Wise}, J.~H., {Eracleous}, M., {Charlton}, J.~C.,  \& {Ganguly}, R.
2004,
  \apj, 613, 129

\bibitem[\protect\citeauthoryear{{Yamamoto} \& {Vansevi{\v c}ius}}{{Yamamoto}
  \& {Vansevi{\v c}ius}}{1999}]{yv99}
{Yamamoto}, T.~M.,  \& {Vansevi{\v c}ius}, V. 1999, \pasj, 51, 405

\bibitem[\protect\citeauthoryear{{York} et~al.}{{York} et~al.}{2000}]{york00}
{York}, D.~G., et~al. 2000, \aj, 120, 1579

\bibitem[\protect\citeauthoryear{{Yuan} \& {Wills}}{{Yuan} \&
  {Wills}}{2003}]{yw03}
{Yuan}, M.~J.,  \& {Wills}, B.~J. 2003, \apjl, 593, L11

\bibitem[\protect\citeauthoryear{{Zhou} et~al.}{{Zhou} et~al.}{2006}]{zhou06}
{Zhou}, H., {Wang}, T., {Wang}, H., {Wang}, J., {Yuan}, W.,  \&
{Lu}, Y. 2006,
  \apj, 639, 716

\end{thebibliography}


\begin{deluxetable}{lccccccrrrccc}
\tablewidth{0pc}
\tabletypesize{\tiny}
\tablecaption{Measured and Derived Quasar Parameters}
\tablehead {
\colhead{Name} &
\colhead{MJD} &
\colhead{Plate} &
\colhead{Fiber} &
\colhead{Redshift} &
\colhead{$F_\lambda$} &
\colhead{$\alpha_\mathrm{NUV}$} &
\colhead{$\lambda L_\lambda$} &
\colhead{Mg\,{\sc ii}} &
\colhead{$M_\mathrm{BH}$} &
\colhead{$L_\mathrm{bol}/$} &
\colhead{Run} &
\colhead{Class} \\
 & & & & & & & &
\colhead{FWHM} & & \colhead{$L_\mathrm{Edd}$}
}
\startdata
SDSS J$000009.42-102751.88$ & 52143 & 0650 & 199 & 1.844230 &  2.2672 & 1.802 &   4.577 & 12320 & 30.856 & 0.0491 & 2 & AAL\\
SDSS J$000058.24-004646.29$ & 51791 & 0387 & 093 & 1.895120 &  2.9567 & 1.763 &   6.496 &  4560 &  4.983 & 0.4316 & 1 & Reg\\
SDSS J$000118.42-010221.60$ & 51791 & 0387 & 042 & 1.968500 &  1.2266 & 1.843 &   3.033 &  9200 & 14.180 & 0.0708 & 3 & AAL\\
SDSS J$000119.64+154828.78$ & 52235 & 0750 & 566 & 1.924210 &  2.3214 & 1.301 &   5.347 &  4885 &  5.219 & 0.3393 & 3 & BAL\\
SDSS J$000139.43+152624.11$ & 52235 & 0750 & 541 & 1.789660 &  5.8318 & 1.246 &  10.727 &  4370 &  5.793 & 0.6131 & 1 & Reg\\
SDSS J$000139.64-103824.05$ & 52143 & 0650 & 136 & 1.820630 &  1.9547 & 1.017 &   3.792 &  4300 &  3.441 & 0.3649 & 2 & Reg\\
SDSS J$000219.34-105259.58$ & 52143 & 0650 & 097 & 1.762000 &  0.9092 & 0.436 &   1.594 &  5245 &  3.406 & 0.1549 & 2 & AAL\\
SDSS J$000221.80+151454.59$ & 52235 & 0750 & 553 & 1.823530 &  3.3027 & 1.123 &   6.438 &  3840 &  3.519 & 0.6058 & 3 & AAL\\
SDSS J$000319.87-090156.19$ & 52143 & 0650 & 561 & 1.760220 &  9.9334 & 1.637 &  17.356 &  3390 &  4.371 & 1.3148 & 3 & Reg\\
SDSS J$000422.30-090219.36$ & 52143 & 0650 & 609 & 1.858810 &  5.1395 & 1.323 &  10.633 &  3370 &  3.431 & 1.0262 & 2 & Reg\\
SDSS J$000510.84-092534.79$ & 52143 & 0650 & 621 & 1.866540 &  1.6132 & 1.170 &   3.381 &  3635 &  2.330 & 0.4805 & 3 & Reg\\
SDSS J$000546.49-002413.98$ & 51793 & 0388 & 268 & 1.729750 &  2.0840 & 1.133 &   3.450 &  3705 &  2.443 & 0.4675 & 3 & Reg\\
SDSS J$000612.30-104310.65$ & 52141 & 0651 & 258 & 1.735310 &  3.3973 & 1.481 &   5.680 &  3455 &  2.686 & 0.7003 & 2 & Reg\\
SDSS J$000629.29-093550.93$ & 52141 & 0651 & 429 & 1.758180 &  5.0479 & 0.936 &   8.788 &  6880 & 13.075 & 0.2226 & 2 & Reg\\
SDSS J$000648.74-094037.00$ & 52141 & 0651 & 468 & 1.835040 &  2.5152 & 0.686 &   5.000 &  3970 &  3.340 & 0.4957 & 2 & AAL\\
SDSS J$000654.11-001533.30$ & 51793 & 0388 & 234 & 1.725170 &  7.5664 & 0.943 &  12.422 &  4650 &  7.028 & 0.5853 & 4 & Reg\\
SDSS J$000735.52-085435.16$ & 52141 & 0651 & 417 & 1.777750 &  2.4386 & 1.561 &   4.394 &  3315 &  2.191 & 0.6639 & 2 & Reg\\
SDSS J$000759.40+150822.65$ & 52251 & 0751 & 557 & 1.968020 &  2.4355 & 0.887 &   6.017 &  2915 &  1.964 & 1.0143 & 2 & Reg\\
SDSS J$000815.33-095854.03$ & 52141 & 0651 & 494 & 1.949790 &  6.6211 & 0.854 &  15.891 &  4420 &  7.129 & 0.7381 & 1 & AAL\\
SDSS J$000904.43-004332.85$ & 51793 & 0388 & 098 & 1.828630 &  3.8336 & 1.089 &   7.538 &  3700 &  3.518 & 0.7094 & 4 & Reg\\
SDSS J$000919.26+152355.17$ & 52251 & 0752 & 343 & 1.934950 &  1.4671 & 1.883 &   3.439 &  4730 &  3.976 & 0.2864 & 2 & AAL\\
SDSS J$001017.52-100238.90$ & 52138 & 0652 & 316 & 1.819020 &  2.8847 & 1.216 &   5.581 &  4075 &  3.705 & 0.4987 & 2 & AAL\\
SDSS J$001017.81+010450.74$ & 51793 & 0388 & 607 & 1.817960 &  3.6331 & 0.992 &   7.016 &  5120 &  6.514 & 0.3566 & 3 & AAL\\
SDSS J$001020.25-103059.09$ & 52141 & 0651 & 107 & 1.742540 &  2.5056 & 1.424 &   4.243 &  5045 &  4.993 & 0.2814 & 3 & AAL\\
SDSS J$001029.92-101145.54$ & 52138 & 0652 & 302 & 1.751640 &  3.2360 & 1.281 &   5.569 &  3665 &  2.994 & 0.6159 & 1 & Reg\\
SDSS J$001053.58+000642.84$ & 51795 & 0389 & 348 & 1.873370 &  2.6721 & 1.169 &   5.664 &  5655 &  7.185 & 0.2610 & 3 & Reg\\
SDSS J$001054.41-085438.66$ & 52141 & 0651 & 569 & 1.745720 &  9.3966 & 1.043 &  16.003 &  4485 &  7.364 & 0.7196 & 4 & Reg\\
SDSS J$001111.14+151034.83$ & 52251 & 0751 & 632 & 1.731630 &  2.7532 & 1.609 &   4.573 &  5700 &  6.602 & 0.2294 & 1 & Reg\\
SDSS J$001142.79+152216.30$ & 52251 & 0752 & 438 & 1.886190 &  3.6573 & 0.219 &   7.918 &  3270 &  2.812 & 0.9323 & 4 & Reg\\
SDSS J$001157.76-103735.21$ & 52138 & 0652 & 264 & 1.857680 &  4.2954 & 0.774 &   8.870 &  4060 &  4.573 & 0.6423 & 4 & AAL\\
SDSS J$001244.62+141112.97$ & 52251 & 0752 & 217 & 1.705560 &  3.2471 & 0.842 &   5.146 &  5660 &  6.881 & 0.2476 & 4 & AAL\\
SDSS J$001330.58+001814.28$ & 51795 & 0389 & 385 & 1.844230 &  1.8880 & 1.713 &   3.812 &  4270 &  3.401 & 0.3711 & 2 & Reg\\
SDSS J$001341.74+143531.31$ & 52251 & 0752 & 194 & 1.933230 &  1.8982 & 1.433 &   4.437 &  6480 &  8.412 & 0.1746 & 3 & AAL\\
SDSS J$001400.45+004255.49$ & 51795 & 0389 & 412 & 1.709720 &  4.5593 & 1.438 &   7.280 &  4045 &  4.137 & 0.5827 & 3 & Reg\\
SDSS J$001408.23-085242.19$ & 52138 & 0652 & 363 & 1.744560 &  4.5117 & 0.137 &   7.668 &  6825 & 12.068 & 0.2104 & 2 & BAL\\
SDSS J$001411.00-084429.15$ & 52138 & 0652 & 375 & 1.767220 &  2.9899 & 0.287 &   5.289 &  5470 &  6.510 & 0.2690 & 2 & AAL\\
SDSS J$001438.28-010750.12$ & 51795 & 0389 & 211 & 1.815640 &  4.8011 & 0.553 &   9.234 &  3295 &  3.070 & 0.9961 & 2 & BAL\\
SDSS J$001453.37-002827.52$ & 51795 & 0389 & 239 & 1.922620 &  3.7010 & 0.566 &   8.503 &  4405 &  5.278 & 0.5335 & 2 & Reg\\
SDSS J$001507.00-000800.81$ & 51795 & 0389 & 196 & 1.703300 & 10.1183 & 0.906 &  15.970 &  3020 &  3.336 & 1.5852 & 1 & Reg\\
SDSS J$001545.33-095754.12$ & 52138 & 0652 & 479 & 1.795910 &  1.3341 & 1.949 &   2.481 &  4780 &  3.483 & 0.2358 & 2 & Reg\\
SDSS J$001657.00+005532.06$ & 51900 & 0390 & 331 & 1.756100 &  3.0960 & 1.595 &   5.370 &  3775 &  3.123 & 0.5694 & 3 & Reg\\
SDSS J$001710.85+135556.52$ & 52251 & 0752 & 044 & 1.808170 & 10.4602 & 1.386 &  19.863 &  4360 &  7.704 & 0.8538 & 1 & AAL\\
SDSS J$001741.85-105613.30$ & 52138 & 0652 & 041 & 1.806080 &  3.6452 & 1.407 &   6.897 &  3555 &  3.115 & 0.7331 & 2 & Reg\\
SDSS J$001800.21+004602.82$ & 51795 & 0389 & 611 & 1.900380 &  3.3131 & 1.584 &   7.342 &  4690 &  5.583 & 0.4354 & 2 & Reg\\
SDSS J$001826.80-091038.78$ & 52138 & 0652 & 578 & 1.857870 &  2.9839 & 0.911 &   6.164 &  4105 &  3.940 & 0.5180 & 2 & Reg\\
SDSS J$001913.57-103848.25$ & 52138 & 0652 & 070 & 1.862490 &  2.2511 & 0.822 &   4.686 &  4245 &  3.704 & 0.4189 & 4 & Reg\\
SDSS J$001959.48-090809.46$ & 52145 & 0653 & 324 & 1.792750 &  1.4842 & 1.450 &   2.745 &  6195 &  6.135 & 0.1481 & 3 & Reg\\
SDSS J$002000.50+155110.02$ & 52233 & 0753 & 374 & 1.751000 &  1.9210 & 1.737 &   3.302 &  4470 &  3.484 & 0.3139 & 3 & AAL\\
SDSS J$002028.35-002915.00$ & 51900 & 0390 & 234 & 1.926880 &  2.7439 & 1.765 &   6.348 &  4165 &  4.112 & 0.5111 & 2 & AAL\\
SDSS J$002028.97+153435.91$ & 52233 & 0753 & 430 & 1.764390 &  4.2464 & 1.156 &   7.474 &  4035 &  4.168 & 0.5938 & 2 & AAL\\
SDSS J$002127.88+010420.39$ & 51900 & 0390 & 443 & 1.819400 &  4.9046 & 1.518 &   9.494 &  7615 & 16.610 & 0.1893 & 2 & BAL\\
SDSS J$002132.63-092424.31$ & 52145 & 0653 & 422 & 1.869800 &  2.8025 & 1.081 &   5.905 &  5690 &  7.419 & 0.2636 & 3 & Reg\\
SDSS J$002143.30+010840.26$ & 51900 & 0390 & 534 & 1.901220 &  2.2417 & 1.618 &   4.974 &  4790 &  4.850 & 0.3396 & 3 & Reg\\
SDSS J$002302.56+153446.39$ & 52233 & 0753 & 462 & 1.723460 &  3.2186 & 1.294 &   5.268 &  4790 &  4.983 & 0.3501 & 2 & BAL\\
SDSS J$002344.36+143115.43$ & 52233 & 0753 & 165 & 1.866000 &  5.4223 & 1.334 &  11.353 &  7715 & 18.544 & 0.2027 & 1 & BAL\\
SDSS J$002411.66-004348.06$ & 51782 & 0391 & 290 & 1.794470 &  6.6288 & 1.839 &  12.295 &  4575 &  6.770 & 0.6014 & 3 & AAL\\
SDSS J$002439.64-091600.58$ & 52145 & 0653 & 538 & 1.792960 &  3.5242 & 1.638 &   6.519 &  4625 &  5.135 & 0.4204 & 3 & Reg\\
SDSS J$002448.07-090055.05$ & 52145 & 0653 & 530 & 1.858650 &  3.2093 & 1.476 &   6.638 &  4510 &  4.924 & 0.4464 & 3 & Reg
\enddata
\tablecomments{[The complete version of this table is in the electronic edition of
the Astrophysical Journal and from http://physics.uwyo.edu/agn/.  The printed edition contains only a sample.] Units
on $F_\lambda$\ are $10^{-17}$\,erg cm$^{-2}$\ s$^{-1}$\ \AA$^{-1}$. Units on $\lambda
L_\lambda$\ are $10^{45}$\,erg s$^{-1}$. Both $F_\lambda$\ and $\lambda L_\lambda$\
are taken at 3000\,\AA. Units on the {Mg\,{\sc ii}} FWHM are \kms. Units on $M_\mathrm{BH}$\ are
$10^8\,M_\odot$.}
\label{tab:qsos}
\end{deluxetable}

\end{document}